\documentclass[nohyper]{JHEP3}

\usepackage{axodraw}
\usepackage{url}
\usepackage{epsfig}
\usepackage{latexsym}
\usepackage{amsmath}
\def\mathswitch#1{\relax\ifmmode#1\else$#1$\fi}

\newcommand{\si}{\sigma}
\newcommand{\sq}{\tilde{q}}
\newcommand{\sqb}{\bar{\tilde{q}}}
\newcommand{\gl}{\tilde{g}}

\newcommand{\NLO}{\mathrm{NLO}}

\newcommand{\ord}{{\cal O}}

\newcommand{\TeV}{\unskip\,\mathrm{TeV}}

\def\nn{\nonumber}

\def\text{\textstyle}

\def\bc{\begin{center}}
\def\ec{\end{center}}
\def\bi{\begin{itemize}}
\def\ei{\end{itemize}}

\title{Soft-gluon resummation for squark and gluino hadroproduction }

\author{Wim Beenakker\\
  Theoretical High Energy Physics, Radboud University Nijmegen, P.O. Box 9010\\
  NL-6500 GL Nijmegen, The Netherlands}
\author{Silja Brensing, Michael Kr\"amer, Anna Kulesza\\
  Institut f\"ur Theoretische Physik, RWTH Aachen University\\
  D-52056 Aachen, Germany}
\author{Eric Laenen\\
  ITFA, University of Amsterdam, Valckenierstraat 65, 1018 XE Amsterdam, \\
  ITF, Utrecht University, Leuvenlaan 4, 3584 CE Utrecht\\
  Nikhef Theory Group, Science Park 105, 1098 XG Amsterdam, The
  Netherlands}
\author{ Irene Niessen\\
  Theoretical High Energy Physics, Radboud
  University Nijmegen, P.O. Box 9010\\
  NL-6500 GL Nijmegen, The Netherlands}

\abstract{We consider the resummation of soft gluon emission for
  squark and gluino hadroproduction at next-to-leading-logarithmic
  (NLL) accuracy in the framework of the minimal supersymmetric
  standard model. We present analytical results for squark-squark and
  squark-gluino production and provide numerical predictions for all
  squark and gluino pair-production processes at the Tevatron and at
  the LHC.  The size of the soft-gluon corrections and the reduction
  in the scale uncertainty are most significant for processes
  involving gluino production. At the LHC, where the sensitivity to
  squark and gluino masses ranges up to 3~TeV, the corrections due to
  NLL resummation over and above the NLO predictions can be as high as
  35\% in the case of gluino-pair production, whereas at the Tevatron,
  the NLL corrections are close to 40\% for squark-gluino final states
  with sparticle masses around 500~GeV.  }


\preprint{ITP-UU-09/33\\
  NIKHEF/2009-015}

\begin{document}

\section{Introduction}
\label{sec:intro}

The search for supersymmetry (SUSY)~\cite{Golfand:1971iw, Wess:1974tw}
is among the most important tasks at current and future colliders.
Squarks and gluinos, the coloured supersymmetric particles, are
expected to be produced most copiously in hadronic collisions.
Searches at the proton--antiproton collider Tevatron with a
centre-of-mass energy of $\sqrt{S}=1.96$~TeV have placed lower limits
on squark and gluino masses in the range of
300-400~GeV~\cite{:2007ww,Aaltonen:2008rv}.  The proton--proton
collider LHC with $\sqrt{S}=14$~TeV design energy will extend the
range of sensitivity to squarks and gluinos with masses up to about
3~TeV~\cite{Aad:2009wy,Bayatian:2006zz,gianotti_eps}.

In the minimal supersymmetric extension of the Standard Model (MSSM)
\cite{Nilles:1983ge, Haber:1984rc} with R-parity conservation, squarks
and gluinos are pair-produced in collisions of two hadrons $h_1$ and
$h_2$:
\begin{equation}
  h_1 h_2 \;\to\; \tilde{q}\tilde{q}\,,
  \tilde{q}\sqb\,, \tilde{q}\tilde{g}\,, \tilde{g}\tilde{g} + X\,.
\label{eq:processes}
\end{equation}
In Eq.~(\ref{eq:processes}) and throughout the rest of this paper we
suppress the chiralities of the squarks $\tilde{q} =(\tilde{q}_{L},
\tilde{q}_{R})$ and do not explicitly state the charge-conjugated
processes.  We include squarks $\sq$ of any flavour except for top
squarks.  The production of top squarks~\cite{Beenakker:1997ut} has to
be considered separately since the strong Yukawa coupling between top
quarks, top squarks and Higgs fields gives rise to potentially large
mixing effects and mass splitting~\cite{Ellis:1983ed}.

Accurate theoretical predictions for inclusive cross sections are
crucial to derive exclusion limits for squark and gluino
masses~\cite{:2007ww,Aaltonen:2008rv} and, in the case of discovery,
can be used to determine sparticle masses~\cite{Baer:2007ya} and
properties~\cite{Kane:2008kw}. The cross sections for the squark and
gluino pair-production processes (\ref{eq:processes}) are known at
next-to-leading order (NLO) in SUSY-QCD~\cite{Beenakker:1994an,
  Beenakker:1995fp, Beenakker:1996ch}. Electroweak corrections to the
$\ord (\alpha_{\rm s}^2)$ tree-level production~\cite{Hollik:2007wf,
  Hollik:2008yi, Hollik:2008vm, Mirabella:2009ap} and the electroweak
Born production channels of $\ord (\alpha\alpha_{\rm s})$ and $\ord
(\alpha^2)$~\cite{Alan:2007rp, Bornhauser:2007bf} are significant for
the pair production of SU(2)-doublet squarks $\tilde{q}_L$ and at
large invariant masses in general, but they are moderate for total
cross sections summed over all squark species.

The NLO SUSY-QCD corrections to squark and gluino hadroproduction
reduce the renormalization- and factorization-scale dependence of the
predictions.  In general these corrections also significantly increase
the cross section with respect to the Born
predictions~\cite{Kane:1982hw, Harrison:1982yi,Dawson:1983fw} if the
renormalization and factorization scales are chosen close to the
average mass of the pair-produced sparticles. A significant part of
these large corrections can be attributed to the threshold region
where the partonic centre-of-mass energy is close to the kinematic
threshold for producing massive particles. In this region the NLO
corrections are dominated by the contributions due to soft gluon
emission off the coloured particles in the initial and final state and
by the Coulomb corrections due to the exchange of gluons between the
massive sparticles in the final state. The soft-gluon corrections can
be taken into account to all orders in perturbation theory by means of
threshold resummation.

Previous work has addressed the soft-gluon resummation for
squark-antisquark and gluino-gluino production at
next-to-leading-logarithmic (NLL) accuracy
\cite{Kulesza:2008jb,Kulesza:2009kq}. For the squark-antisquark
production process the dominant contribution to the
next-to-next-to-leading order (NNLO) correction coming from the
resummed cross section at next-to-next-to-leading-logarithmic (NNLL)
level has been studied in~\cite{Langenfeld:2009eg}. Moreover, a
formalism allowing for the resummation of soft and Coulomb gluons in
the production of coloured sparticles has been presented
in~\cite{Beneke:2009rj,Beneke:2009nr}, and bound state effects have
been studied for gluino-pair production in
Ref.~\cite{Hagiwara:2009hq}.  Additionally, threshold resummation for
single colour-octet scalar production at the LHC has been investigated
in~\cite{Idilbi:2009cc}.

In this work, we present the analytical components needed to perform
NLL resummation for squark-squark and squark-gluino pair-production.
In addition, we provide numerical predictions for the entire set
(\ref{eq:processes}) of pair-production processes of coloured
sparticles at the Tevatron and the LHC.

The paper is structured as follows. In section~\ref{se:resummation} we
review the formalism of soft-gluon resummation. The calculation of the
one-loop soft anomalous dimension matrices for the $\sq \sq$ and $\sq
\gl$ production processes is discussed in
section~\ref{se:anomalous_dim}. We present numerical results for
squark and gluino production at the Tevatron and the LHC in
section~\ref{se:numres} and conclude in section~\ref{se:conclusion}.
A more detailed description of certain aspects of our calculation and
some explicit formulae that enter the expressions for the resummed
cross sections are collected in the appendices.

\section{Soft-gluon resummation}
\label{se:resummation}

In this section we review the formalism of threshold resummation for
the production of a pair of coloured massive particles. Since the
corresponding theoretical expressions have already been discussed in
detail in Ref.~\cite{Kulesza:2009kq}, we shall be brief.

The inclusive hadroproduction cross section $\si_{h_1h_2\rightarrow
  kl}$ for two massive SUSY particles $k$ and $l$, where $k,l$ can be
a squark ($\tilde{q}$), antisquark ($\bar{\tilde{q}}$) or gluino
($\tilde{g}$), can be written in terms of its partonic version
$\si_{ij\rightarrow kl}$ as
\begin{multline}
  \label{eq:7}
  \si_{h_1 h_2 \to k l}\bigl(\rho, \{m^2\}\bigr) 
  \;=\; \sum_{i,j} \int d x_1 d x_2\,d\hat{\rho}\;
        \delta\left(\hat{\rho} - \frac{\rho}{x_1 x_2}\right)\\
        \times\,f_{i/h_{1}}(x_1,\mu^2 )\,f_{j/h_{2}}(x_2,\mu^2 )\,
        \si_{ij \to kl}\bigl(\hat{\rho},\{ m^2\},\mu^2\bigr)\,,
\end{multline}
where $\{m^2\}$ denotes all masses entering the calculations, $i,j$
are the initial parton flavours, $f_{i/h_1}$ and $f_{j/h_2}$ the
parton distribution functions, and $\mu$ is the common factorization
and renormalization scale. The hadronic threshold for inclusive
production of two final-state particles with masses $m_3$ and $m_4$
corresponds to a hadronic center-of-mass energy squared that is equal
to $S=(m_3+m_4)^2$.  Thus we define the threshold variable $\rho$,
measuring the distance from threshold in terms of energy fraction, as
$$
\rho \;=\; \frac{(m_3+m_4)^2}{S}\,.
$$
The partonic equivalent of this threshold variable is defined as
$\hat{\rho}=\rho/(x_1x_2)$, where $x_{1,2}$ are the momentum fractions
of the partons. This is a generalized version of the threshold
variable used e.g.\ in Ref.~\cite{Kulesza:2009kq}. It accounts for
unequal masses of the pair-produced particles in the final state,
making it applicable to the case of squark-gluino production.

In the threshold region, the most dominant contributions to the
higher-order QCD corrections due to soft gluon emission have the
general form\footnote{See section~\ref{se:anomalous_dim} for more
  discussion on the form of a threshold variable in the case of
  unequal masses.}
\begin{equation}
\alpha_{\rm s}^n \log^m\!\beta^2\ \ , \ \ m\leq 2n 
\qquad {\rm \ with\ } \qquad 
\beta^2 \,\equiv\, 1-\hat{\rho} \,=\, 1 \,-\, \frac{(m_3+m_4)^2}{s}\,,
\label{eq:logbeta:structure}
\end{equation}
where $s=x_1x_2S$ is the partonic center-of-mass energy squared.  The
resummation of the soft-gluon contributions is performed after taking
a Mellin transform (indicated by a tilde) of the cross section,
\begin{eqnarray}
  \label{eq:10}
  \tilde\si_{h_1 h_2 \to kl}\bigl(N, \{m^2\}\bigr) 
  &\equiv& \int_0^1 d\rho\;\rho^{N-1}\;
           \si_{h_1 h_2\to kl}\bigl(\rho,\{ m^2\}\bigr) \\[2mm] \nn
  &=&      \;\sum_{i,j} \,\tilde f_{i/{h_1}} (N+1,\mu^2)\,
           \tilde f_{j/{h_2}} (N+1, \mu^2) \,
           \tilde{\si}_{ij \to kl}\bigl(N,\{m^2\},\mu^2\bigr)\,.
\end{eqnarray}
The logarithmically enhanced terms are then of the form $\alpha_{\rm
  s}^n \log^m N$, $m\leq 2n$, with the threshold limit
$\beta\rightarrow 0$ corresponding to $N\rightarrow \infty$.  The
resummed cross section takes the schematic form \cite{Sdy,CTdy}
\begin{equation}
  \label{eq:3}
  \tilde\si_{h_1 h_2 \to kl} (N) 
  \;=\; \exp\Big[L g_1(\alpha_{\rm s}L) + g_2(\alpha_{\rm s}L) + \ldots \Big]  
        \times P(\alpha_{\rm s})\,,
\end{equation}
in which all dependence on the large logarithm $L=\log N$ occurs in
the exponent, and no term in the perturbative series $P(\alpha_{\rm
  s})$ grows with increasing $N$. Keeping only the $g_1$ term
constitutes the leading logarithmic (LL) approximation, including also
the $g_2$ term is called the next-to-leading logarithmic (NLL)
approximation, etc. Up to NLL accuracy it suffices to keep the
lowest-order term in $P$.

The all-order summation of such logarithmic terms depends on the
near-threshold factorization of the cross sections into functions that
each capture the effects of classes of radiation effects: hard,
collinear (including soft-collinear), and wide-angle soft radiation
\cite{Sdy,CTdy,Contopanagos:1996nh,
  Kidonakis:1998bk,Kidonakis:1998nf,Bonciani:1998vc}
\begin{eqnarray}
  \label{eq:11}
  \tilde \si_{ij  \to k l}\bigl(N, \{m^2\},\mu^2\bigr) 
  &=& \Delta_i (N+1,Q^2,\mu^2)\,\Delta_j (N+1,Q^2,\mu^2) \nn\\[2mm]
  &&  \times\,\sum_{IJ} H_{ij\rightarrow k l, JI}\bigl(N,\{m^2\},\mu^2\bigr)
      \,\bar{S}_{ij\rightarrow k l ,IJ}\bigl(Q/(N\mu),\mu^2\bigr)\,,
\end{eqnarray}
where we have introduced the hard scale $Q^2 = (m_3+m_4)^2$.  Before
we comment on each function separately, we recall that soft radiation
is coherently sensitive to the colour structure of the hard process
from which it is emitted
\cite{Botts:1989kf,Contopanagos:1996nh,Kidonakis:1997gm,Kidonakis:1998bk}.
The various structures are labelled by the indices $I,J$ in a way made
more precise further below.

The functions $\Delta_{i}$ and $\Delta_{j}$ sum the effects of the
(soft-)collinear radiation from the incoming partons.  They are
process-independent and do not depend on the colour structures.  They
contain the leading logarithmic dependence, as well as part of the
subleading logarithmic behaviour, and are listed e.g.\ in
Ref.~\cite{Kulesza:2009kq}.

The function $H_{ij\rightarrow k l ,JI}$ incorporates only
higher-order effects of hard, off-shell partons and therefore does not
contain $\log N$ dependence. This hard function depends on the colour
representations of the external particles in the partonic process.
There are usually multiple tensors $c_I$ that can connect these colour
representations, where $I\,$ labels the possible tensors. For
instance, in the case of squark-antisquark (with colour indices
$a_3,a_4$) production by the annihilation of light quarks (with colour
indices $a_1,a_2$) there are two colour tensors, which may be chosen
as
\begin{eqnarray} \label{eq:4} 
c_1(a_1,a_2;a_3,a_4) & = & \delta_{a_1 a_2}\,\delta_{a_3 a_4}\qquad 
(s{\scriptscriptstyle -}\mathrm{channel\;\; singlet}),\nonumber\\[2mm]
c_2(a_1,a_2;a_3,a_4) & = & T^c_{a_1 a_2}T^c_{a_3 a_4}\qquad
  (s{\scriptscriptstyle -}\mathrm{channel\;\; octet}) \,.
\end{eqnarray}
The hard function $H_{ij\rightarrow k l ,JI}$ is a matrix in this
colour-tensor space, with the indices $JI$ indicating the colour
structure.  Note that we paired the indices in example (\ref{eq:4})
according to the $s$-channel. Other choices are possible as well
\cite{Botts:1989kf, Kidonakis:1997gm,Kidonakis:1998nf}, but choosing
an $s$-channel basis will be convenient at threshold.

The soft function $\bar{S}_{ij\rightarrow k l ,IJ}$ in
Eq.~(\ref{eq:11}) is also a matrix in colour-tensor space, since soft
emissions mix the connecting colour tensors. This soft function is
constructed \cite{Kidonakis:1997gm,Kidonakis:1998nf} from an eikonal
cross section, which in turn is defined in terms of the square of
expectation values of products of Wilson-line operators belonging to
the external particles in the process. These Wilson lines generate to
all orders the soft-gluon radiation in the process and depend on the
direction and colour representation of the corresponding external
particle.  To avoid double counting with the $\Delta_{i}$ and
$\Delta_{j}$ factors in Eq.~(\ref{eq:11}), the expectation values are
divided by the square of expectation values of the Wilson lines
themselves.  In this way, collinear-soft radiation already included in
the $\Delta_{i}$ and $\Delta_{j}$ factors is removed. What remains is
a soft function whose perturbation series takes the form $\alpha_{\rm
  s}^n \log^m\!N$, $m\leq n$, and therefore contributes only at NLL
accuracy.

Although the combination of the soft and collinear functions in the
cross section is gauge invariant, the functions themselves are not
automatically separately gauge invariant. The collinear functions only
depend on the colour representations of the incoming partons.
Therefore the gauge dependence of the soft function cannot depend on
the colour structure of the process either. This implies that we can
make the soft and collinear functions separately gauge invariant by
rescaling them with a scalar in colour-tensor space. This rescaling
has implicitly been performed in Eq.~(\ref{eq:11}), where the soft
function has been divided by $\sqrt{S^{\rm sing}_{i\bar{i}}}
\sqrt{S^{\rm sing}_{j\bar{j}}}$\, as indicated by the bar on
$\bar{S}_{ij\to kl, IJ}$. The factor $S^{\rm sing}_{i\bar{i}}$ is the
soft function for two incoming Wilson lines of flavour $i$ and $\bar
i$ annihilating into a colour-singlet\footnote{Note that if the colour
  representations are $\mathbf{3}$ and $\bar{\mathbf{3}}$ this
  corresponds to the Drell-Yan process. For octets, it corresponds to
  Higgs production by gluon fusion.}.  By taking the square root of
such a soft function, we effectively isolate the gauge dependence of a
single line. Therefore this procedure works not only for $q\bar{q}$ or
$gg$ initial states but also for initial states that cannot annihilate
into a colour-singlet, such as $qg$ and $qq$.  To compensate for the
division factor in the soft function, the collinear functions
$\Delta_{i}$ and $\Delta_{j}$ have been multiplied by the factors
$\sqrt{S^{\rm sing}_{i\bar{i}}}$\, and $\sqrt{S^{\rm
    sing}_{j\bar{j}}}$\, respectively. Analytical expressions for
these functions given in the literature (see
e.g.~Ref.~\cite{Kulesza:2009kq}) explicitly include this
multiplicative factor.

Near threshold the soft function reduces considerably. For the
inclusive cross section and our choice of colour basis, the matrix
$\bar{S}_{ij\to kl, IJ}$ becomes diagonal in colour-tensor space in
the threshold limit $\beta\rightarrow 0$ \cite{Kulesza:2009kq}. In
this limit we have (suppressing particle flavour labels)
\begin{equation}
  \lim_{\beta \to 0}\,  \bar{S}_{IJ}\bigl(Q/(N\mu), \mu^2\bigr) 
  \;=\; \delta_{IJ}\,S^{(0)}_{IJ}\,\Delta_{I}^{\rm (s)}
        \bigl(Q/(N\mu),\mu^2\bigr)
\end{equation}
with
\begin{equation}
  \Delta_{I}^{\rm (s)}\bigl(Q/(N\mu),\mu^2\bigr) 
  \;=\; \exp\Big[\int_{\mu}^{Q/N}\frac{dq}{q}\,\frac{\alpha_{\rm s}(q)}{\pi}
                 \,D_{I} \,\Big]\,,
\label{eq:1}
\end{equation}
where $S^{(0)}_{IJ}$ is the lowest-order expression for the soft
function, given by
\begin{equation}
  \label{eq:9}
  {S}^{(0)}_{IJ} \;=\; {\rm{tr}} \left( c^{\dagger}_I\, c_J \right) \,.
\end{equation}
The one-loop coefficients $D_{I}$ are defined by
\begin{equation}
  D_{I} \;\equiv\; \lim_{\beta\to 0}\, \frac{\pi}{\alpha_{\mathrm s}}\,
                   2\,{\rm Re}\,(\bar \Gamma_{II})\,.
\label{eq:2}
\end{equation}
The values of the $D_{I}$ coefficients for $\sq\sq$ and $\sq \gl$
production are calculated in section~\ref{sec:boldm-thresh-limit}.
The form of Eq.~(\ref{eq:1}) follows from a renormalization-group
equation for $\bar{S}_{IJ}(Q/(N\mu))$
\cite{Contopanagos:1996nh,Kidonakis:1998nf}, with one-loop anomalous
dimensions $\bar{\Gamma}_{ij\to kl, IJ}$, often referred to as the
``soft'' anomalous-dimension matrix. If the calculations are performed
in the axial-gauge with gauge vector $n^\mu$, the one-loop anomalous
dimensions are given by
\begin{equation}
  \label{eq:5}
  \bar\Gamma_{IJ} 
  \;=\;  
  \Gamma_{IJ}\,-\, \frac{\alpha_{\rm s}}{2\pi}\,
  \sum_{p=\{i,j\}} C_{2,p}\left(1-\log\Bigl(2\,\frac{(v_p\cdot n)^2}{|n|^2}\,
  \Bigr)-i\pi\right)\delta_{IJ}\,,
\end{equation}
where the sum is over the two incoming particles, and $|n|^2 = -n^2 -
i\epsilon$, see Ref.~\cite{Botts:1989kf}. The dimensionless vector $v_p$ is
given by the momentum of the incoming massless particle $p$ multiplied
by $\sqrt{2/s}$. The factors $C_{2,p}$ are either $C_F$ or $C_A$,
depending on whether $p$ is a quark or gluon, respectively. The
subtraction exhibited in Eq.~(\ref{eq:5}) results from the division by
the factor $\sqrt{S^{\rm sing}_{i\bar{i}}} \sqrt{S^{\rm
    sing}_{j\bar{j}}}$\, described before. The matrix $\Gamma_{IJ}$ is
the anomalous dimension matrix of the products of Wilson-line
operators connected by the various possible colour tensors mentioned
earlier. More details on its calculation are given in
section~\ref{sec:soft-anom-dimens}.

In the threshold limit the resummed partonic cross section becomes
\begin{eqnarray}
  \label{eq:12}
  \tilde{\sigma}^{\rm (res)} _{ij\rightarrow kl}\bigl(N,\{m^2\},\mu^2\bigr) 
  &=& \sum_{I}\,
      \tilde\sigma^{(0)}_{ij\rightarrow kl,I}\bigl(N,\{m^2\},\mu^2\bigr)\, 
      C_{ij \rightarrow kl, I}\bigl(N,\{m^2\},\mu^2\bigr) \\[1mm]
  && \times\,\Delta_i (N+1,Q^2,\mu^2)\,\Delta_j (N+1,Q^2,\mu^2)\,
     \Delta^{\rm (s)}_{ij\rightarrow kl,I}\bigl(Q/(N\mu),\mu^2\bigr)\,, \nn
\end{eqnarray}
where $\tilde{\sigma}^{(0)}_{ij \rightarrow kl, I}$ are the
leading-order (LO) cross sections in Mellin-moment space. For the case
of $\sq\sq$ and $\sq \gl$ production we present them in
appendix~\ref{sec:appA}. The functions $C_{ij \rightarrow kl, I}$ are
of perturbative nature and contain information about hard
contributions beyond leading order.  This information is only relevant
beyond NLL accuracy and therefore we keep $C_{ij \rightarrow kl,I} =1
$ in our calculations.

Having constructed the NLL cross-section in the Mellin-moment space,
the inverse Mellin transform has to be performed in order to recover
the hadronic cross section $\si_{h_1 h_2 \to kl}$. In order to retain
the information contained in the NLO cross
sections~\cite{Beenakker:1994an, Beenakker:1995fp,Beenakker:1996ch},
the NLO and NLL results are combined through a matching procedure that
avoids double counting of the logarithmic terms in the following way:
\begin{eqnarray}
  \label{eq:14}
  \si^{\rm (NLL+NLO~matched)}_{h_1 h_2 \to kl}\bigl(\rho, \{m^2\},\mu^2\bigr) 
  \;&=&\; \si^{\rm (NLO)}_{h_1 h_2 \to kl}\bigl(\rho, \{m^2\},\mu^2\bigr)
          \\[1mm]
    && \hspace*{-3ex} +\, \sum_{i,j=q,\bar{q},g}\,\int_\mathrm{CT}\,\rho^{-N}\,
       \tilde f_{i/h_1}(N+1,\mu^2)\,\tilde f_{j/h_{2}}(N+1,\mu^2) \nn\\[3mm]
    && \hspace*{-2ex}\times\,
       \left[\tilde\si^{\rm(res)}_{ij\to kl}\bigl(N,\{m^2\},\mu^2\bigr)
             \,-\, \tilde\si^{\rm(res)}_{ij\to kl}\bigl(N,\{m^2\},\mu^2\bigr)
       {\left.\right|}_{\scriptscriptstyle({\NLO})}\, \right]. \nn
\end{eqnarray}
We adopt the ``minimal prescription'' of Ref.~\cite{Catani:1996yz} for
the contour CT of the inverse Mellin transform in Eq.~(\ref{eq:14}).
In order to use standard parametrizations of parton distribution
functions in $x$-space we employ the method introduced in
Ref.~\cite{Kulesza:2002rh}.

\section{\boldmath Soft anomalous dimensions and Born cross sections for $\sq\sq$ and $\sq\gl$ 
                   production}
\label{se:anomalous_dim}

\subsection{Kinematics}\label{s:kinematics}

To set the stage for the discussion of the soft anomalous dimensions
we first introduce the relevant kinematical definitions that are used
in the calculation. We consider the following generic process
\begin{equation}
i(a_1, p_1) \,j(a_2,p_2) \;\to\ k(a_3, p_3) \,l(a_4,p_4)\,,
\label{eq:generic:proc}
\end{equation}
where the colour indices $a_i$ and the momenta of the particles $p_i$
are given in parentheses. In those cases where a final-state squark
features in the process, summation over both squark chiralities
($\tilde{q}_{_L}$ and $\tilde{q}_{_R}$) and all possible squark
flavours is implied, the latter being restricted by the choice of
initial-state quark flavours.  For the processes investigated here,
i.e.~squark-squark ($kl= \sq\sq$) and squark-gluino ($kl=\sq\gl$)
production, top-squark final states are not possible since top quarks
are excluded as initial-state partons. In view of the absence of
top-squark final states, all squark-flavour and chirality states are
considered to be mass degenerate with mass $\,m_{\tilde q}$. The
gluino mass is denoted by $\,m_{\tilde g}$.

All analytical results presented in section~\ref{se:anomalous_dim} are
derived for a general SU$(N_{_C})$-theory, with $N_{_C}$ the number of
colours.  This means that the colour indices $a_i$ for gluons and
gluinos can take $N_{_C}^2-1$ different values, since these particles
are in the adjoint representation. For (s)quarks, which are in the
fundamental representation, the colour indices are $N_{_C}$-valued.

The particle momenta featuring in the generic process
(\ref{eq:generic:proc}) obey the on-shell conditions
$\,p_1^2=p_2^2=0\,$, $\,p_3^2=m_3^2\;$ and $\;p_4^2=m_4^2$.  For the
kinematical description of the reactions the standard Mandelstam
invariants
\begin{equation}
s \,=\, (p_1+p_2)^2
\quad,\quad
t \,=\, (p_1-p_3)^2
\qquad\mbox{and}\qquad
u \,=\, (p_1-p_4)^2
\label{eq:Mandelstam}
\end{equation}
are used.  In the centre-of-mass frame of the final-state particles
the absolute value of the final-state momenta can then be written as
\begin{equation}
\label{eq:6}
|\vec{p}_3|_{\mbox{\tiny cm}} \ =\ |\vec{p}_4|_{\mbox{\tiny cm}}
\ =\ \frac{1}{2}\,\kappa\beta\sqrt{s}\,,
\end{equation}
with $\beta=\sqrt{1-(m_3+m_4)^2/s}$ defined in Eq.~(\ref{eq:logbeta:structure})
and
\begin{equation}
\kappa \ \equiv\ \sqrt{1-\frac{(m_3-m_4)^2}{s}}\,.
\label{eq:kappa}
\end{equation}

The presence of the factor $\kappa$ is special to the case of unequal
masses. As Eq.~(\ref{eq:6}) shows, it occurs quite naturally in
matrix-element expressions for the processes we consider in this
study. We could have defined the variable $\beta'=\kappa\beta$ and
taken moments with respect to this variable. Instead we have opted to
use the variable $\beta$ in our calculations in order to facilitate
convolutions underlying the resummation. Because $\log\beta' =
\log\beta + \log\kappa$, choosing $\beta'$ would have resulted in
different subleading logarithmic terms.  To NLL accuracy these
differences in the expressions for the resummed partonic cross
sections are cancelled by different terms arising from the
convolutions.

In order to present the results for the leading-order partonic cross
sections it is helpful to introduce two more shorthand notations:
\begin{equation}
m_+^2\ \equiv\ m_{\tilde g}^2+m_{\tilde q}^2
\qquad\mbox{and}\qquad
m_-^2\ \equiv\ m_{\tilde g}^2-m_{\tilde q}^2\,.
\label{eq:mplusmminus}
\end{equation}

\subsection{Colour bases in the $s$-channel}\label{s:colour}

As discussed in section~\ref{se:resummation}, colour correlations need
to be taken into account once NLL soft-gluon resummation is performed
for processes involving pair-production of coloured particles.  To
this end an appropriate colour basis has to be chosen. We have opted
to use an $s$-channel colour basis, which traces the colour flow
through the $s$-channel and has the virtue of rendering the anomalous
dimension matrices diagonal at threshold \cite{Kidonakis:1997gm,
  Kulesza:2008jb, Kulesza:2009kq, Beneke:2009rj}.

Since we are dealing with two coloured particles in both initial and
final state, the $s$-channel basis is obtained by performing an
$s$-channel colour decomposition of the reducible two-particle product
representations into irreducible ones. For squark-squark and
squark-gluino production this amounts to the following decompositions
in SU$(3)$:
\begin{align}
qq\to\tilde q\tilde q:&\qquad 
\mathbf{3}\otimes\mathbf{3} \;=\; \mathbf{\bar 3}\oplus\mathbf{6}\nonumber\,, 
\\[2mm]
qg\to\tilde q\tilde g:&\qquad 
\mathbf{3}\otimes\mathbf{8} \;=\; \mathbf{3}\oplus\mathbf{\bar 6}
                                  \oplus\mathbf{15}\,,
\label{eq:irreps}
\end{align}
where the product representations apply to both the initial and final
state.  In a general SU$(N_{_C})$-theory the dimensions of the various
representations are of course different, but the number of base
tensors for these two processes remains the same.

An economic way to construct the $s$-channel colour bases for
squark-squark and squark-gluino production is to start with an
arbitrary complete colour basis of the considered process in terms of
which the $s$-channel base tensors $c_I(a_1,a_2;a_3,a_4)$ can be
expressed. Then the $s$-channel basis can be obtained by simply
requiring that a particular base tensor is orthogonal to all other
base tensors and projects on itself when contracted in $s$-channel:
\begin{align}\label{eq:selfprojective}
\sum_{b,b'}\,c_I(a_1,a_2;b,b')\,c_{I'}(b,b';a_3,a_4)
\;=\; Z\delta_{II'}\,c_I(a_1,a_2;a_3,a_4)\,, 
\end{align}
where $Z$ is an arbitrary normalization constant. A similar procedure
was found by the authors of Ref.~\cite{Beneke:2009rj} on the basis of
an analysis in terms of Clebsch--Gordon coefficients.  This projective
construction of the $s$-channel base tensors constitutes a direct way
of obtaining explicit implementations of the irreducible
representations on the right-hand side of Eq.~(\ref{eq:irreps}).  The
minimal requirement for the projective method to work is that the
particles in the initial state must be in the same representations as
those in the final state, as follows directly from the fact that the
labels of the initial state are contracted with those of the final
state in Eq.~(\ref{eq:selfprojective}).  This is indeed the case for
both the squark-squark and squark-gluino production processes. An
example of the calculation of the $s$-channel colour basis for the
$qq\to\tilde q\tilde q$ process is given in appendix~\ref{sec:appB}.
 
In order to present the $s$-channel base tensors in the subsequent
text, we will need the following SU$(N_{_C})$-objects: the singlet
colour structures $\delta_{ab}$, where $a$ and $b$ belong to particles
in either the adjoint or the fundamental representation, the
generators of the fundamental representation $T^c_{ab}$, the structure
constants $f_{abc}$ and the symmetric forms $d_{abc}$.

\subsection{Leading-order partonic cross sections}
\label{sec:loxs_colour}

Having defined all necessary ingredients, we can now present the
results for the colour-decomposed $\sq\sq$ and $\sq\gl$ partonic cross
sections at LO. These partonic cross sections are averaged over
initial-state spin and colour. The colour-decomposed LO cross sections
for the $q\bar{q} \to \sq\sqb\,$, $gg \to \sq\sqb\,$, $q\bar{q} \to
\gl\gl\,$ and $\,gg \to \gl\gl$ processes, together with their
Mellin-moment transforms, can be found in Ref.~\cite{Kulesza:2009kq}.

\subsubsection{Squark-squark production}
We consider the process
\begin{equation}
q_{f_1}(a_1,p_1)\,q_{f_2}(a_2,p_2)
\;\to\ \tilde q(a_3,p_3)\,\tilde q(a_4,p_4)\,,
\end{equation}
where the flavours of the initial-state quarks are indicated by
$f_{1}$, $f_2$ and all external particles are in the fundamental
representation of SU$(N_{_C})$.  The method described in
section~\ref{s:colour} to obtain a suitable $s$-channel colour basis
yields the following two colour tensors:
\begin{equation}
c_1^{qq} \ =\ \delta_{a_1a_4}\delta_{a_2a_3}-\delta_{a_1a_3}\delta_{a_2a_4}
\qquad\mbox{and}\qquad
c_2^{qq} \ =\ \delta_{a_1a_4}\delta_{a_2a_3}+\delta_{a_1a_3}\delta_{a_2a_4}\,.
\label{eq:qqtoqqbasis}
\end{equation}
The dimensions of the representations spanned by these two base
tensors are given by $\,\mbox{dim}(R_1^{qq}) =
\frac{1}{2}N_{_C}(N_{_C}-1)\;$ and $\;\mbox{dim}(R_2^{qq}) =
\frac{1}{2}N_{_C}(N_{_C}+1)$. In the SU$(3)\,$ case this basis
coincides up to normalization factors with the base tensors given in
Ref.~\cite{Beneke:2009rj} for the $\,\mathbf{\bar 3}\,$ and
$\,\mathbf{6}\,$ representations. The decomposition of the LO partonic
squark-pair cross section in terms of the base tensors
(\ref{eq:qqtoqqbasis}) is given by
\begin{eqnarray*}
\sigma^{(0)}_{qq\to\tilde{q}\tilde{q},1} &=&
\frac{\pi\alpha_{\rm s}^2(N_{_C}^2-1)(N_{_C}+1)}{4N_{_C}^3s}\,
\biggl[\, \frac{2m_{\tilde g}^2}{2m_-^2 + s}\,L_1\,\delta_{f_1f_2}
    \,-\, \frac{2m_-^2+s}{s}\,L_1 
    \,-\, \frac{2m_-^4+ sm_{\tilde g}^2}{m_-^4+sm_{\tilde g}^2}\,\beta\,
\biggr]\,,\\[3mm]
\sigma^{(0)}_{qq\to\tilde{q}\tilde{q},2} &=& 
\frac{\pi\alpha_{\rm s}^2(N_{_C}^2-1)(N_{_C}-1)}{4N_{_C}^3s}\,
\biggl[\, \frac{-\,2m_{\tilde g}^2}{2m_-^2 + s}\,L_1\,\delta_{f_1f_2}
    \,-\, \frac{2m_-^2+s}{s}\,L_1
    \,-\, \frac{2m_-^4+ sm_{\tilde g}^2}{m_-^4+sm_{\tilde g}^2}\,\beta\,
\biggr]\,,
\end{eqnarray*}
with 
\begin{equation*}
L_1 \ \equiv\ \log\left(\frac{s+2m_-^2-s\beta}{s+2m_-^2+s\beta}\right)\,.
\end{equation*}
The quantities $\,\beta\,$ and $\,m_-^2\,$ are defined in
Eqs.~(\ref{eq:logbeta:structure}) and (\ref{eq:mplusmminus}), using
$\,m_3=m_4=m_{\tilde q}$. The occurrence of the Kronecker-delta
$\,\delta_{f_1f_2}\,$ reflects the fact that for equal-flavoured
initial-state quarks extra diagrams contribute.  In
appendix~\ref{sec:appA} we present results for the Mellin-moment
transforms of these colour-decomposed LO cross sections.

\subsubsection{Squark-gluino production}
At the partonic level the $\sq\gl$ production process is given by
\begin{equation}
 q_{f_1}(a_1,p_1)\,g(a_2,p_2) \;\to\ \tilde q(a_3,p_3)\,\tilde g(a_4,p_4)\,.
\end{equation}
The initial and final state of this process involves both a particle
in the fundamental representation ($q$ or $\tilde{q}$) and a particle
in the adjoint representation \mbox{($g$ or $\tilde{g}$)}. For the
\mbox{$s$-channel} colour decomposition the following three base
tensors are used:
\begin{align}
&c_1^{qg} \ =\ \bigl(T^{a_4}T^{a_2}\bigr)_{a_3a_1}\,,\nonumber\\[3mm]
&c_2^{qg} \ =\ \frac{N_{_C}-2}{N_{_C}}\,\delta_{a_2a_4}\delta_{a_1a_3}
\,-\, 2d_{ca_4a_2}T^c_{a_3a_1}
\,+\, 2\,\frac{N_{_C}-2}{N_{_C}-1}\,(T^{a_4}T^{a_2})_{a_3a_1}\,,
\nonumber\\[1mm]
&c_3^{qg} \ =\ \frac{N_{_C}+2}{N_{_C}}\,\delta_{a_2a_4}\delta_{a_1a_3}
\,+\, 2d_{ca_4a_2}T^c_{a_3a_1}
\,-\, 2\,\frac{N_{_C}+2}{N_{_C}+1}\,(T^{a_4}T^{a_2})_{a_3a_1}\,.
\label{eq:qgtoqgbasis}
\end{align}
The dimensions of the representations spanned by these three base
tensors are given by $\mbox{dim}(R_1^{qg}) = N_{_C}$,
$\,\mbox{dim}(R_2^{qg}) = \frac{1}{2}N_{_C}(N_{_C}+1)(N_{_C}-2)\;$ and
\mbox{$\;\mbox{dim}(R_3^{qg}) =
  \frac{1}{2}N_{_C}(N_{_C}\!-\!1)(N_{_C}\!+2)$}.  In the SU$(3)\,$
case this basis coincides up to normalization factors with the base
tensors given in Ref.~\cite{Beneke:2009rj} for the $\,\mathbf{3}$,
$\,\mathbf{\bar 6}\,$ and $\,\mathbf{15}\,$ representations. The
decomposition of the LO partonic squark-gluino cross section in terms
of the base tensors (\ref{eq:qgtoqgbasis}) is given by
\begin{eqnarray*}
\sigma^{(0)}_{qg\to\tilde{q}\tilde{g},1} &=& 
\frac{\alpha_{\rm s}^2\pi}{(N_{_C}^2-1)s}\, 
\biggl[\, \Bigl(\, \frac{2m_{\tilde g}^2m_-^2}{s^2} 
            \,-\, \frac{2m_-^4+s^2+2m_-^2s}{2s^2}\,N_{_C}^2 \Bigr)\,L_2 \\
&& +\; \frac{m_-^2}{s}\,\Bigl(\, \frac{m_-^2-s}{sN_{_C}^2} 
                               + \frac{2m_{\tilde q}^2}{s} \,\Bigr)\,L_3
   \,- \Bigl(\, \frac{7m_-^2+3s}{4s}\,N_{_C}^2 
        -\frac{3m_-^2+s}{2s}+\frac{7m_-^2-s}{4N_{_C}^2s}\,\Bigr)\,\kappa\beta\,
\biggr]\,,\\[2mm]
\sigma^{(0)}_{qg\to\tilde{q}\tilde{g},2} &=&
\frac{\alpha_{\rm s}^2\pi(N_{_C}-2)}{(N_{_C}-1)s}\,
\biggl[\, \frac{2m_-^2(m_+^2-s)-s^2}{4s^2}\,L_2
    \,+\, \frac{m_-^2(m_+^2-s)}{2s^2}\,L_3 \,-\, \frac{m_-^2}{s}\,\kappa\beta\,
\biggr]\,,\\[3mm]
\sigma^{(0)}_{qg\to\tilde{q}\tilde{g},3} &=&
\frac{\alpha_{\rm s}^2\pi(N_{_C}+2)}{(N_{_C}+1)s}\, 
\biggl[\, \frac{2m_-^2(m_+^2-s)-s^2}{4s^2}\,L_2
    \,+\, \frac{m_-^2(m_+^2-s)}{2s^2}\,L_3 \,-\, \frac{m_-^2}{s}\,\kappa\beta\,
\biggr]\,,
\end{eqnarray*}
with 
\begin{equation*}
L_2 \ =\ \log\left(\frac{s+m_-^2-\kappa s\beta}
                        {s+m_-^2+\kappa s\beta}\right)
\qquad\mbox{and}\qquad
L_3 \ =\ \log\left(\frac{s-m_-^2-\kappa s\beta}
                        {s-m_-^2+\kappa s\beta}\right)\,.
\end{equation*}
The quantities $\,\beta\,,\,\kappa\,$ and $\,m_{\pm}^2\,$ are defined
in Eqs.~(\ref{eq:logbeta:structure}), (\ref{eq:kappa}) and
(\ref{eq:mplusmminus}), using $\,m_3=m_{\tilde q}$ and
$\,m_4=m_{\tilde g}$.  In appendix~\ref{sec:appA} we present results
for the Mellin-moment transforms of these colour-decomposed LO cross
sections.

\subsection{The soft anomalous-dimension matrices}
\label{sec:soft-anom-dimens}
As we reviewed in section \ref{se:resummation} below Eq.~(\ref{eq:5}),
resummation to NLL accuracy requires the anomalous dimensions
$\Gamma_{IJ}$ of the products of Wilson-line operators connected by a
base tensor $c_I$.  To this end one must compute the UV divergences
from their loop corrections, and from these the renormalization
constants $Z_{IJ}$ for these operators. Here we only need the one-loop
corrections. The anomalous dimensions can be computed from the
residues of the UV poles in the renormalization constants $Z_{IJ}$ as
\begin{equation}
  \label{eq:8}
  \Gamma_{IJ} \;=\; -\,\alpha_{\rm s} \frac{\partial}{\partial \alpha_{\rm s}} 
                    \,\mathrm{Res}_{\epsilon \rightarrow 0}\, 
                    Z_{IJ}(\alpha_{\rm s},\epsilon)\,.
\end{equation}
The relevant UV divergences occur in loop corrections to the base
tensors $c_I$ \cite{Kidonakis:1997gm,Kidonakis:1998nf} due to the
Wilson lines. The complete first order correction to $c_I$ can be
written as
\begin{equation}
  \label{eq:13}
  \sum_{ij}\,\omega^{ij}{\cal{C}}^{ij}_{IJ}\,c_J\,,
\end{equation}
where $i$ and $j$ denote the eikonal lines between which the gluon is
spanned, $\omega^{ij}$ is the corresponding kinematic part of the
one-loop correction, and ${\cal{C}}^{ij}_{IJ}$ denotes how the base
tensors get mixed due to the corrections.  At one-loop we can
calculate the anomalous dimensions directly from Eq.~(\ref{eq:13})
\begin{equation}
\label{eq:15}
 \Gamma_{IJ} \;=\; -\sum_{ij}\,{\cal{C}}^{ij}_{IJ}\, 
                   {\rm Res_{\epsilon\rightarrow 0}}\,\omega^{ij}\,.
\end{equation}
The precise form of this function depends on the colour basis chosen.
The eikonal integrals that constitute the $\omega^{ij}$ can be found
in Ref.~\cite{Kidonakis:1997gm}, except for the unequal-mass case that
we need for squark-gluino production. The corresponding integral
$\omega^{34}$ is discussed in appendix~\ref{sec:appD}, using the
Feynman rules in the eikonal approximation presented in
appendix~\ref{sec:appC}.
 
In order to present the results for the soft anomalous dimensions in a
compact way, we introduce the following $t$- and $u$-channel
quantities
\begin{align}
& \Lambda \ \equiv\ \frac{1}{2}\,\bigl[ T(m_3)+T(m_4)+U(m_3)+U(m_4) \bigr]\,,
  \nonumber\\[2mm]
& \Omega \ \equiv\ \frac{1}{2}\,\bigl[ T(m_3)+T(m_4)-U(m_3)-U(m_4) \bigr]\,,
\end{align}
in terms of the $t$- and $u$-channel logarithms%
\footnote{Note that in the case of equal masses $m_3=m_4$ the
  quantities $\Lambda, \Omega, T(m)$ and $U(m)$ reduce to the
  corresponding quantities $\bar\Lambda, \bar\Omega, \bar T$ and $\bar
  U$ defined in Ref.~\cite{Kulesza:2009kq}.}
\begin{equation}
T(m) \ =\ \log\left(\frac{m^2-t}{\sqrt{sm^2}}\right)
          \,-\, \frac{1-i\pi}{2}
\qquad\mbox{and}\qquad
U(m) \ =\ \log\left(\frac{m^2-u}{\sqrt{sm^2}}\right)
          \,-\, \frac{1-i\pi}{2}\,. \vspace*{3mm}
\end{equation}

The one-loop soft anomalous-dimension matrices for the $q\bar{q}\to
\sq\sqb\,$, $gg \to \sq\sqb\,$, $q\bar{q}\to \gl\gl$ and $gg \to
\gl\gl$ processes have been calculated in Ref.~\cite{Kulesza:2009kq},
where the corresponding values of the $D_{ij \to kl, I}$ coefficients
can be found as well.\footnote{Note that Ref.~\cite{Kulesza:2009kq}
  uses a subtraction term different from Eq.~(\ref{eq:5}).}

\subsubsection{Soft anomalous dimensions for squark-pair production at
  one-loop}

In the basis (\ref{eq:qqtoqqbasis}) the one-loop soft
anomalous-dimension matrix is given by
\begin{equation}
\bar\Gamma_{qq\to\tilde q\tilde q} \ =\ \frac{\alpha_{\rm s}}{2\pi}
\left( \begin{array}{cc}
       \displaystyle C_2(R_1^{qq})\,\Lambda
       \,-\, \frac{N_{_C}+1}{N_{_C}}\,(L_\beta +1) 
       & -\,(N_{_C}+1)\,\Omega \\[3mm]
       -\,(N_{_C}-1)\,\Omega & \displaystyle C_2(R_2^{qq})\,\Lambda 
                               \,+\, \frac{N_{_C}-1}{N_{_C}}\,(L_\beta
                               +1)
       \end{array} \right)\,,
\end{equation}  
with \begin{equation*}
L_\beta \ =\ \frac{1+\beta^2}{2\beta}\,
             \biggl[\, \log\Bigl(\,\frac{1-\beta}{1+\beta}\,\Bigr) 
                       + i\pi \,\biggr]\,.
\end{equation*}
The coefficients $\,C_2(R_I^{qq})\,$ for $\,I=1,2\,$ are the quadratic
Casimir invariants belonging to the representations spanned by the
base tensors $\,c_I^{qq}$:
\begin{equation}
\label{eq:C2qqtoqq}
C_2(R_1^{qq}) \ =\ \frac{(N_{_C}+1)(N_{_C}-2)}{N_{_C}}
\qquad\mbox{and}\qquad
C_2(R_2^{qq}) \ =\ \frac{(N_{_C}-1)(N_{_C}+2)}{N_{_C}}\,. \vspace*{3mm}
\end{equation}

\subsubsection{Soft anomalous dimensions for squark-gluino production
  at one-loop}
In the basis (\ref{eq:qgtoqgbasis}) the one-loop soft anomalous-dimension 
matrix is given by
\begin{equation}
\bar\Gamma_{qg\to\tilde q\tilde g} 
\ =\ \frac{\alpha_{\rm s}}{2\pi}\left( 
\begin{array}{ccc}
  \displaystyle\bar\Gamma_{11,\,qg}\  
  & \ \displaystyle\frac{4N_{_C}^2(N_{_C}-2)}{(N_{_C}^2-1)(N_{_C}-1)}\,\Omega\ 
  & \ \displaystyle\frac{4N_{_C}^2(N_{_C}+2)}{(N_{_C}^2-1)(N_{_C}+1)}\,\Omega 
  \\[3mm]
  \frac{1}{2}\,\Omega\ 
  & \ \displaystyle\bar\Gamma_{22,\,qg}\  
  & \ \displaystyle\frac{N_{_C}(N_{_C}+2)}{N_{_C}+1}\,\Omega  \\[3mm]
  \frac{1}{2}\,\Omega\                   
  & \ \displaystyle\frac{N_{_C}(N_{_C}-2)}{N_{_C}-1}\,\Omega\ 
  & \ \displaystyle\bar\Gamma_{33,\,qg}  
\end{array} \right)\,,
\end{equation}  
with 
\begin{align}
&\bar\Gamma_{11,\,qg} \ =\ 
 C_2(R_1^{qg})\,\Lambda \,+\, \bigl[C_F+\frac{1}{C_F}\bigr]\,\Omega
 \,-\, \frac{N_{_C}^2+1}{2N_{_C}}\,\bigl[ T(m_{\tilde q})-T(m_{\tilde g})\bigr]
 \,-\, N_{_C}\,(L_{v_3,v_4}+1)\,, \nonumber\\[2mm]
&\bar\Gamma_{22,\,qg}  \ =\
 C_2(R_2^{qg})\,\Lambda \,+\, \bigl[C_F-\frac{1}{N_{_C}-1}\bigr]\,\Omega
 \,-\, \frac{N_{_C}^2+1}{2N_{_C}}\,\bigl[ T(m_{\tilde q})-T(m_{\tilde g})\bigr]
 \,-\, (L_{v_3,v_4}+1)\,, \nonumber\\[2mm]
&\bar\Gamma_{33,\,qg}  \ =\
 C_2(R_3^{qg})\,\Lambda \,+\, \bigl[C_F-\frac{1}{N_{_C}+1}\bigr]\,\Omega 
 \,-\, \frac{N_{_C}^2+1}{2N_{_C}}\,\bigl[ T(m_{\tilde q})-T(m_{\tilde g})\bigr]
 \,+\, (L_{v_3,v_4}+1)\,, \label{eq:Gammadiagqgtoqg}
\end{align}
where 
\begin{equation}
L_{v_3,v_4} =\ \frac{\kappa^2+\beta^2}{2\kappa\beta}\,
                 \biggl[\,\log\Bigl(\,\frac{\kappa-\beta}{\kappa+\beta}\,\Bigr)
                        + i\pi\,\biggr]\,.
\label{eq:Lv3v4} 
\end{equation}
The explicit derivation of Eq.~(\ref{eq:Lv3v4}) is presented in
appendix~\ref{sec:appD}.  The coefficients $\,C_2(R_I^{qg})\,$ for
$\,I=1,2,3\,$ are the quadratic Casimir invariants belonging to the
representations spanned by the base tensors $\,c_I^{qg}$:
\begin{eqnarray}
\label{eq:C2qgtoqg}
&& C_2(R_1^{qg}) \ =\ \frac{N_{_C}^2-1}{2N_{_C}} \ \equiv\ C_F\,,
   \qquad
   C_2(R_2^{qg}) \ =\ \frac{(N_{_C}-1)(3N_{_C}+1)}{2N_{_C}} \nonumber\\[3mm]
&& \mbox{and}\qquad
   C_2(R_3^{qg}) \ =\ \frac{(N_{_C}+1)(3N_{_C}-1)}{2N_{_C}}\,. \vspace*{5mm}
\end{eqnarray}

\subsubsection{\boldmath The threshold limit}
\label{sec:boldm-thresh-limit}
At the production threshold, where $\,\beta\to 0$, the soft
anomalous-dimension matrices become diagonal by virtue of using an
$s$-channel basis.  In addition, the diagonal components become
proportional to the total colour charge of the heavy-particle pair
produced at threshold:
\begin{equation}
D_{ij\to kl,I} = -C_2(R^{ij}_I)\,,
\label{eq:Dcoefficients}
\end{equation}
with $\,C_2(R^{ij}_I)\,$ as given in equation (\ref{eq:C2qqtoqq}) for
squark-pair production and in equation (\ref{eq:C2qgtoqg}) for
squark-gluino production. In the SU$(3)$ case the $D_{ij\to kl,I}$
coefficients for squark-pair production are given by
\[
\{D_{qq\to \sq\sq,I}\} \;=\; \{-4/3,-10/3\}\,,
\]
while for the squark-gluino production process they are
\[
\{D_{qg\to \sq\gl,I}\} \;=\; \{-4/3,-10/3,-16/3\}\,.
\]

\section{Numerical results}
\label{se:numres}

In this section we present numerical results for the NLL-resummed
cross sections matched with the complete NLO results for squark and
gluino pair-production at both the Tevatron ($\sqrt{S}=1.96$~TeV) and
the LHC ($\sqrt{S}=14$~TeV). The matching is performed according to
Eq.~(\ref{eq:14}).  From now on we refer to the matched cross sections
as NLL+NLO cross sections.  We also compare the NLL+NLO predictions
with the corresponding NLO results.  The NLO cross sections are
calculated using the publicly available {\tt PROSPINO}
code~\cite{prospino}, based on the calculations presented in
Refs.~\cite{Beenakker:1994an,Beenakker:1995fp,Beenakker:1996ch}.  As
described in detail in Ref.~\cite{Beenakker:1996ch}, the QCD coupling
$\alpha_{\rm s}$ and the parton distribution functions at NLO are
defined in the $\overline{\rm MS}$ scheme with five active flavours.
The masses of squarks and gluinos are renormalized in the on-shell
scheme, and the SUSY particles are decoupled from the running of
$\alpha_{\rm s}$ and the parton distribution functions. As already
discussed in previous sections, no top-squark final states are
considered.  We sum over squarks with both chiralities
($\tilde{q}_{L}$ and $\tilde{q}_{R}$), which are taken as mass
degenerate, and include the charge-conjugated processes in the
numerical predictions. For convenience we define the average mass of
the sparticle pair $m \equiv (m_3+m_4)/2$, which reduces to the squark
and gluino mass for $\sq \sqb\,$, $\sq\sq$ and $\tilde{g}\tilde{g}$
final states, respectively. The renormalization and factorization
scales $\mu$ are taken to be equal. In order to evaluate hadronic
cross sections we use the 2008 NLO MSTW parton distribution
functions~\cite{Martin:2009iq} with the corresponding $\alpha_{\rm
  s}(M_{Z}^2) = 0.120$. The numerical results have been obtained with
two independent computer codes.

We first discuss the scale dependence of the NLL+NLO matched cross
section for the separate processes $p\bar{p} \to \sq\sqb\,, \sq\sq\,,
\sq\gl\,, \gl\gl + X$ at the Tevatron. Figure~\ref{fig:scale_tev}
shows the NLO and NLL+NLO cross sections for
$m_{\sq}=m_{\gl}=m=500$~GeV as a function of the renormalization and
factorization scale $\mu$.  The value of $\mu$ is varied around the
central scale $\mu_0 = m$ from $\mu=\mu_0/10$ up to $\mu=5 \,\mu_0$.
As anticipated, we observe a reduction of the scale dependence when
going from NLO to NLL+NLO, in particular for $\gl\gl$ and $\sq\gl$
production (Figs.~\ref{fig:scale_tev}b and \ref{fig:scale_tev}d,
respectively). In the case of squark pair-production, on the other
hand, the scale reduction due to soft-gluon resummation is moderate
(see Figs.~\ref{fig:scale_tev}a and \ref{fig:scale_tev}c). We note
that the gluino-pair production cross section
(Fig.~\ref{fig:scale_tev}b) is rather small for this particular choice
of masses because of a suppression of the LO $q\bar{q}\to \gl\gl$
amplitude proportional to $m_{\gl}^2-m_{\sq}^2$ near threshold
(cf.~Eq.~(55) of Ref.~\cite{Beenakker:1996ch}).

At the central scale $\mu=\mu_0=m$ the cross-section predictions are
in general enhanced by soft-gluon resummation.  The relative
$K$-factor $K_{\rm NLL}-1 \equiv \sigma_{\rm NLL+NLO}/\sigma_{\rm
  NLO}-1$ at the Tevatron is displayed in Fig.~\ref{fig:k_tev} for
squark and gluino masses in the range between 200~GeV and 600~GeV. We
show results for various mass ratios $r \equiv
m_{\tilde{g}}/m_{\tilde{q}}$. The soft-gluon corrections are moderate
for $\sq\sqb$ production (Fig.~\ref{fig:k_tev}a), but reach values up
to 27\%, 29\% and 60\% for $\gl\gl\,$, $\sq\sq$ and $\sq\gl$ final
states, respectively, in the range of $r$ we consider. Because of the
increasing importance of the threshold region, the corrections in
general become larger for increasing sparticle masses.  The strong
$r$-dependence of $K_{\rm NLL}$ for gluino-pair production in
Fig.~\ref{fig:k_tev}b is driven by the $r$-dependence of the NLO cross
sections for $q\bar{q}\to \gl\gl$. The large effect of soft-gluon
resummation for $\tilde{q}\tilde{g}$ and $\tilde{g}\tilde{g}$
production can be mostly attributed to the importance of gluon initial
states for these processes. Furthermore, the presence of gluinos in
the final state results in enhancement of the NLL
contributions~\cite{Kulesza:2009kq}, since in this case the Casimir
invariants that enter Eq.~(\ref{eq:1}) reach higher values than for
processes involving only squarks.  The substantial value of $K_{\rm
  NLL}$ for $\sq\sq$ production at the Tevatron is a consequence of
the behaviour of the corresponding NLO corrections, which strongly
decrease with increasing squark mass \cite{Beenakker:1996ch}.

We now turn to the discussion of pair production of squarks and
gluinos at the LHC, i.e.~$pp \to \sq\sqb\,, \sq\sq\,, \sq\gl\,, \gl\gl
+ X$. The results for the processes $pp \to \sq\sqb$ and $pp \to
\gl\gl$ agree with those presented in Refs.~\cite{Kulesza:2008jb,
  Kulesza:2009kq}, while the predictions for $pp \to \sq\sq$ and $pp
\to \sq\gl$ are new.  In Fig.~\ref{fig:scale_lhc} the cross sections
are shown for squark and gluino masses $m_{\sq}=m_{\gl}=m=1$~TeV as a
function of the common renormalization and factorization scale $\mu$.
The scale uncertainty of the theoretical prediction is reduced at
NLL+NLO. Similarly to the Tevatron case, soft-gluon resummation is
most significant for gluino-pair production and squark-gluino
production. For those processes, the relative $K$-factor $K_{\rm NLL}
-1$ reaches 35\% for gluino-pair production and 18\% for squark-gluino
production at the highest accessible sparticle masses around 3~TeV
(see Figs.~\ref{fig:k_lhc}b and \ref{fig:k_lhc}d).  The $r$-dependence
of $K_{\rm NLL}$ for gluino-pair production is again driven by the
$r$-dependence of the NLO cross section, discussed in
Ref.~\cite{Beenakker:1996ch}.

Representative values for the NLO and NLL+NLO cross sections at the
Tevatron and the LHC are collected in Tables~\ref{tab:tev} and
\ref{tab:lhc} for equal squark and gluino masses.

The impact of the NLL resummation on the cross section for inclusive
squark and gluino production, i.e.~$p\bar{p}/pp \to
\sq\sqb+\sq\sq+\sq\gl+\gl\gl+X$, can be inferred from the inclusive
$K$-factor displayed in Fig.~\ref{fig:k_inclusive}. The pattern
exhibited in Fig.~\ref{fig:k_inclusive} can be understood from the
relative importance of the $\sq\sqb\,,\sq\sq\,, \sq\gl$ and $\gl\gl$
final states and from their individual \mbox{$K$-factors} as shown in
Figs.~\ref{fig:k_tev} and \ref{fig:k_lhc}.  At $m_{\tilde{q}} =
m_{\tilde{g}} \approx 400$~GeV, for example, the inclusive cross
section at the Tevatron (Fig.~\ref{fig:k_inclusive}a) is built up from
the individual final states in the ratio $\sq\sq : \gl \gl : \sq\gl :
\sq\sqb \approx 1 : 3.6 : 14 : 32$, as can be read off from
Table~\ref{tab:tev}. Owing to the large NLL corrections for the
$\sq\gl$ final state, the resulting inclusive $K$-factor $K_{\rm NLL}$
is approximately~1.1. At $m_{\tilde{q}} = m_{\tilde{g}} = 600$~GeV the
correction to the inclusive cross section at the Tevatron due to NLL
resummation can be as high as 18\%. The inclusive corrections are
smaller at the LHC for sparticle masses below 3~TeV (see
Fig.~\ref{fig:k_inclusive}b). Given the sparticle mass ranges that we
consider, this is consistent with the fact that the distance from
threshold, i.e.~the value of the variable $1-\rho=1-4m^2/S$, is on
average larger at the LHC than at the Tevatron.

In Figs.~\ref{fig:total:matched}a and \ref{fig:total:matched}b we show
for the Tevatron and LHC, respectively, the resummed NLL+NLO total
cross section for inclusive squark and gluino production as a function
of the average sparticle mass $m$. For illustration we show these
results for the choice $m_{\sq} = m_{\gl}$. The error bands indicate
the theoretical uncertainty of the NLL+NLO total cross section due to
the scale variation in the range $m/2\le\mu\le 2m$. The results
presented in Fig.~\ref{fig:total:matched} are the most accurate
theoretical predictions currently available for the above processes.
The reduction of the theoretical error due to variation of the common
factorization and renormalization scale $\mu$ between $\mu = m/2$ and
$\mu = 2m$ is illustrated in Fig.~\ref{fig:total:scale}a for the
Tevatron and in Fig.~\ref{fig:total:scale}b for the LHC. Both at the
Tevatron and at the LHC, soft-gluon resummation leads to a significant
reduction in this part of the theoretical uncertainty.

\section{Conclusions}
\label{se:conclusion}

We have performed the NLL resummation of soft gluon emission for
squark and gluino hadroproduction. Explicit analytical results are
presented for the anomalous dimension matrices and the
colour-decomposed LO cross sections in $x$ and $N$-space for the
$\sq\sq$ and $\sq\gl$ final states. We provide NLO+NLL matched
numerical predictions for all pair-production processes of coloured
sparticles at the Tevatron and the LHC. The NLL corrections lead to a
significant reduction of the scale dependence and, in general,
increase the NLO cross sections. The effect of soft-gluon resummation
is most pronounced for processes with initial-state gluons and
final-state gluinos, which involve a large colour charge.
Specifically, at the Tevatron we find an increase of the cross-section
prediction of up to 40\% at sparticle masses around 500~GeV when going
from NLO to NLL+NLO, depending in detail on the final state and the
ratio of squark to gluino masses. For the inclusive sparticle cross
section at the Tevatron, summed over all pair-production processes for
squarks and gluinos, the enhancement can be as large as approximately
15\% in the mass range up to $500$~GeV, probed by current experimental
searches.  At the LHC, the NLL corrections are particularly
significant for squark-gluino production and gluino-pair production,
reaching approximately 20\% and 30\%, respectively, for sparticle
masses around 3~TeV. Both at the Tevatron and at the LHC, the
inclusion of NLL corrections leads to a reduction of the scale
dependence over the full mass range that will be probed by
experiments. In addition, the NLL corrections lead to a significant
enhancement of the NLO cross-section predictions for heavy sparticles.
The NLL+NLO matched predictions presented in this paper should thus be
used to interpret current and future searches for supersymmetry at the
Tevatron and the LHC.

\section*{Acknowledgments}
This work has been supported in part by the Helmholtz Alliance
``Physics at the Terascale'', the DFG Graduiertenkolleg ``Elementary
Particle Physics at the TeV Scale'', the Foundation for Fundamental
Research of Matter (FOM), the National Organization for Scientific
Research (NWO), the DFG SFB/TR9 ``Computational Particle Physics'',
and the European Community's Marie-Curie Research Training Network
under contract MRTN-CT-2006-035505 ``Tools and Precision Calculations
for Physics Discoveries at Colliders''. MK and EL would like to thank
the CERN TH division for their hospitality.

\clearpage

\FIGURE{
\hspace{-1.0cm}
\begin{tabular}{ll}
(a)\epsfig{file=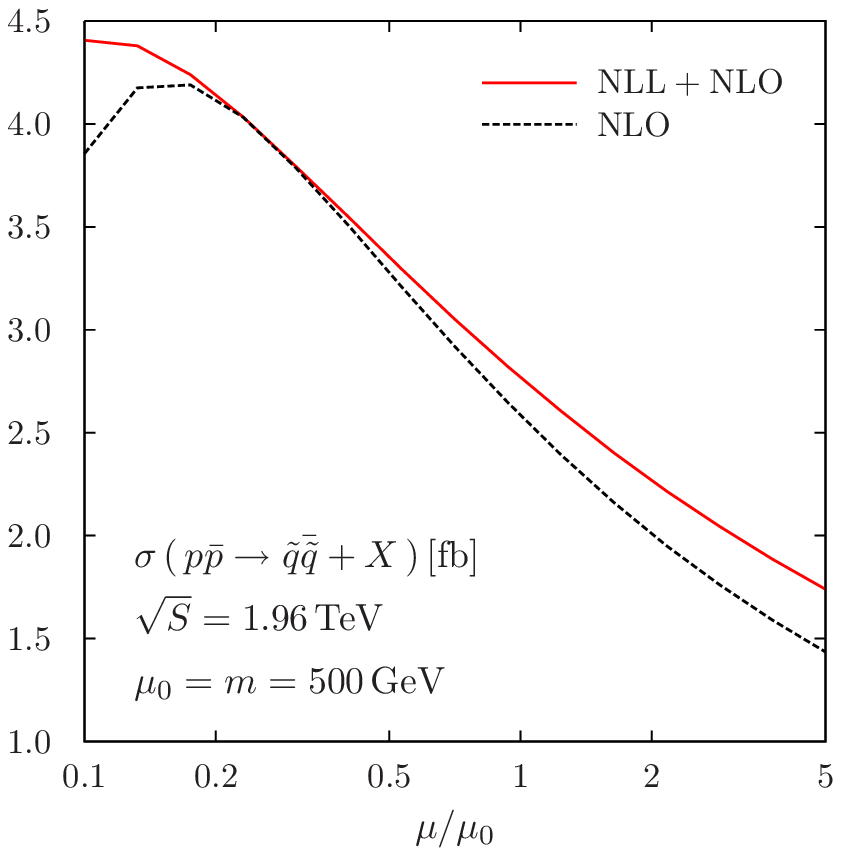, width=0.465\columnwidth}& 
(b)\epsfig{file=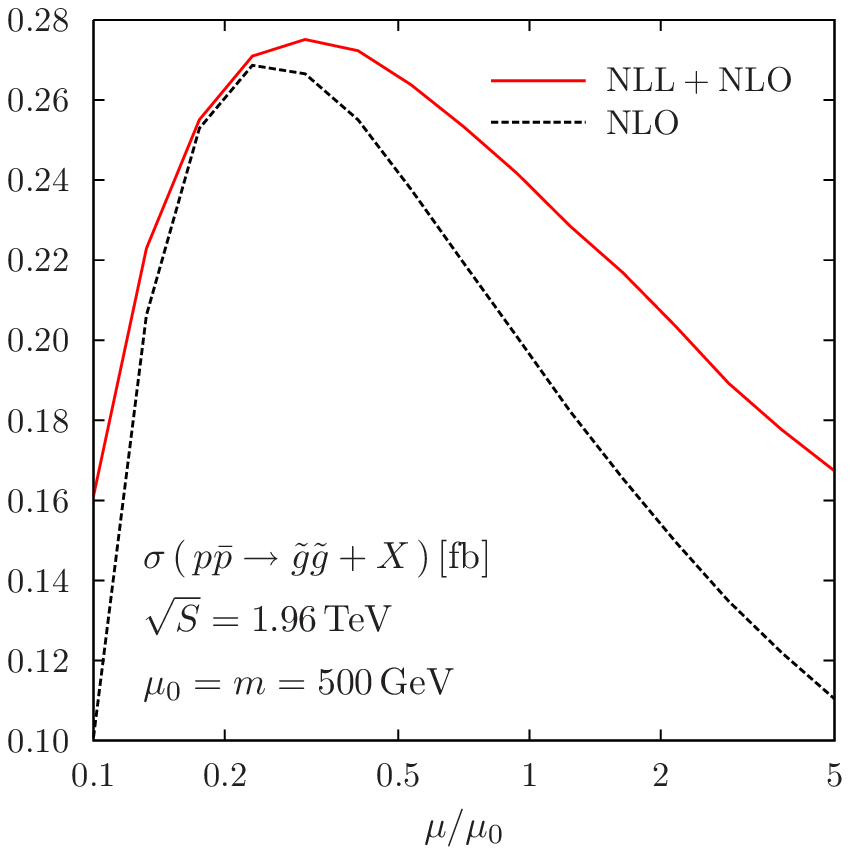, width=0.465\columnwidth }\\
(c)\epsfig{file=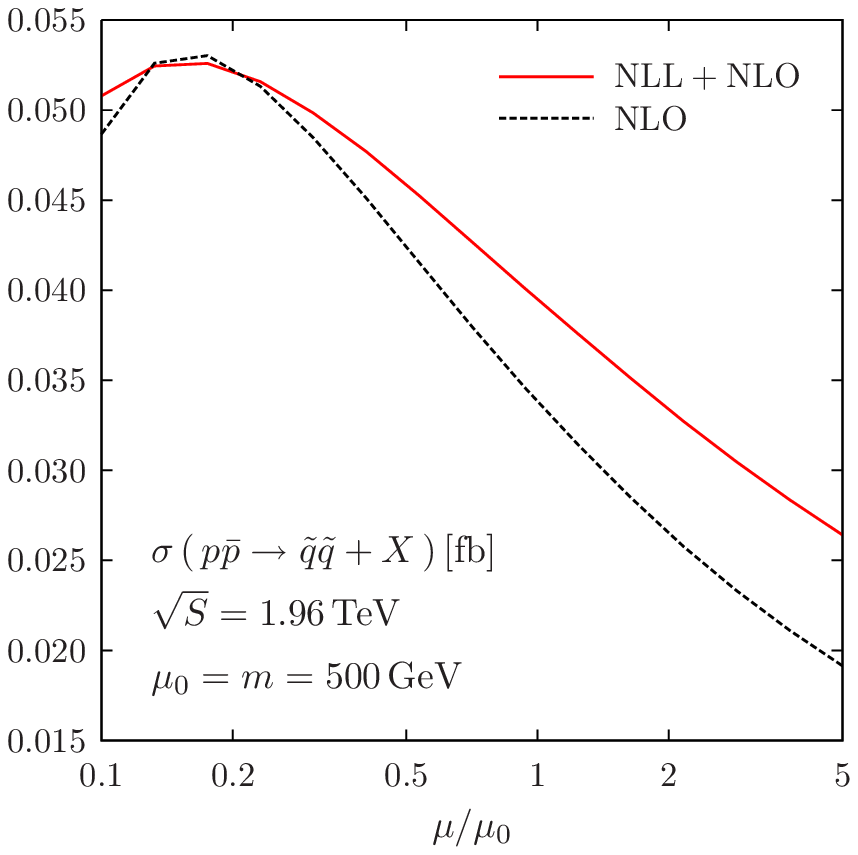, width=0.465\columnwidth}& 
(d)\epsfig{file=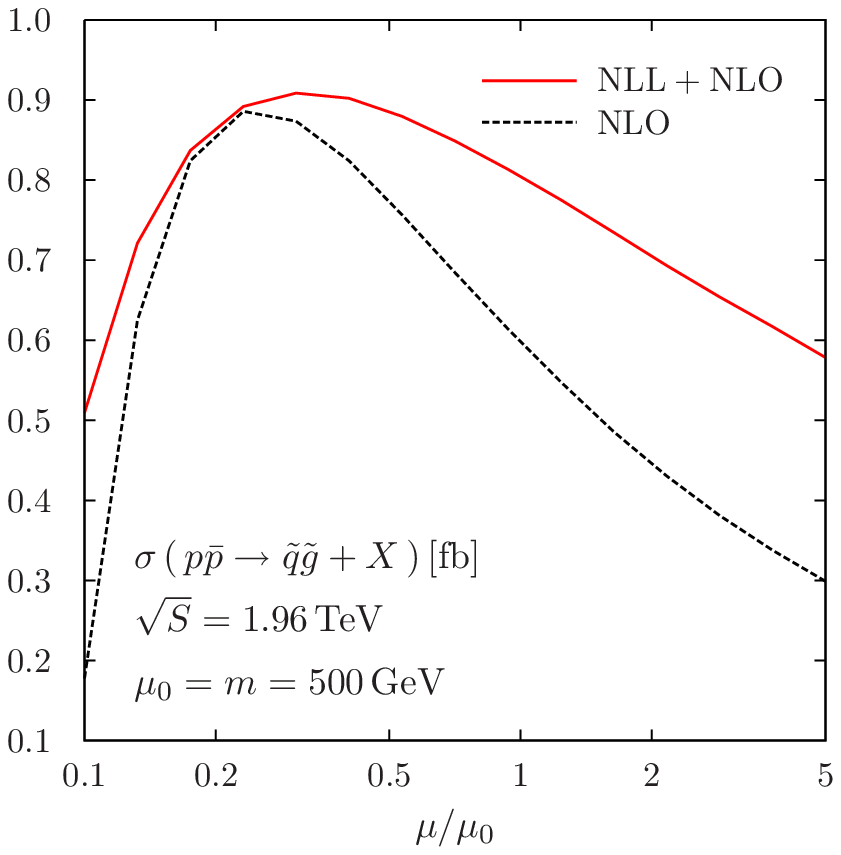, width=0.465\columnwidth }\\
\end{tabular}
\caption{The scale dependence of the NLL+NLO and the NLO total cross
  sections for squark and gluino pair-production processes at the
  Tevatron. The squark and gluino masses have been set to
  $m_{\sq}=m_{\gl}=m=500$~GeV.}
\label{fig:scale_tev}
}

 \FIGURE{
\hspace{-1.0cm}
 \begin{tabular}{ll}
(a)\epsfig{file=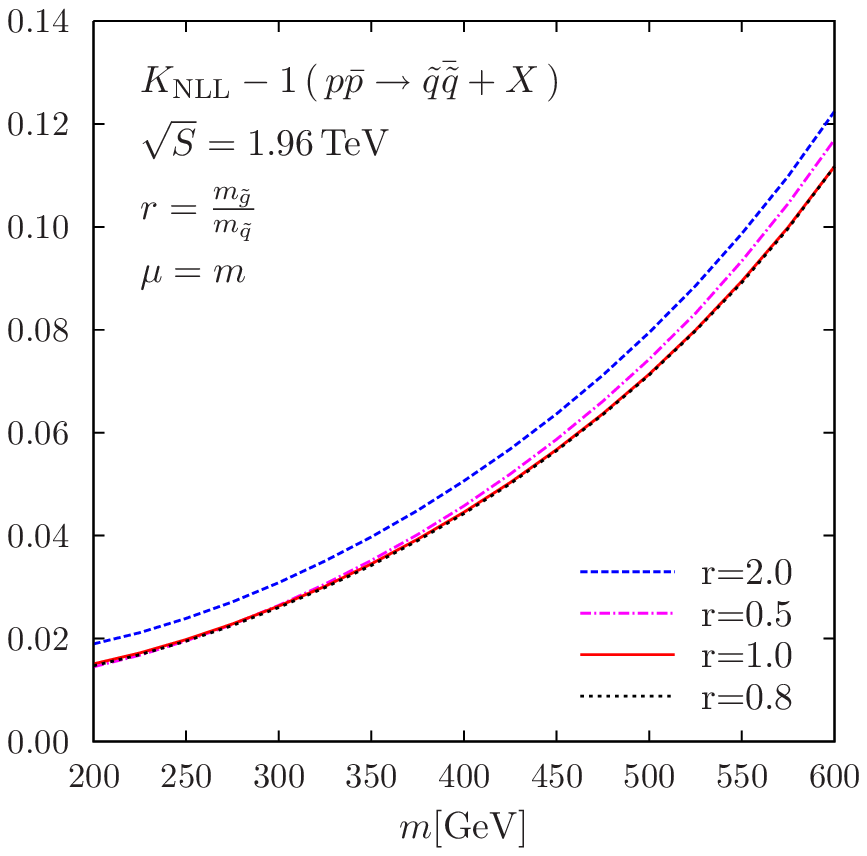, width=0.465\columnwidth} & 
(b)\epsfig{file=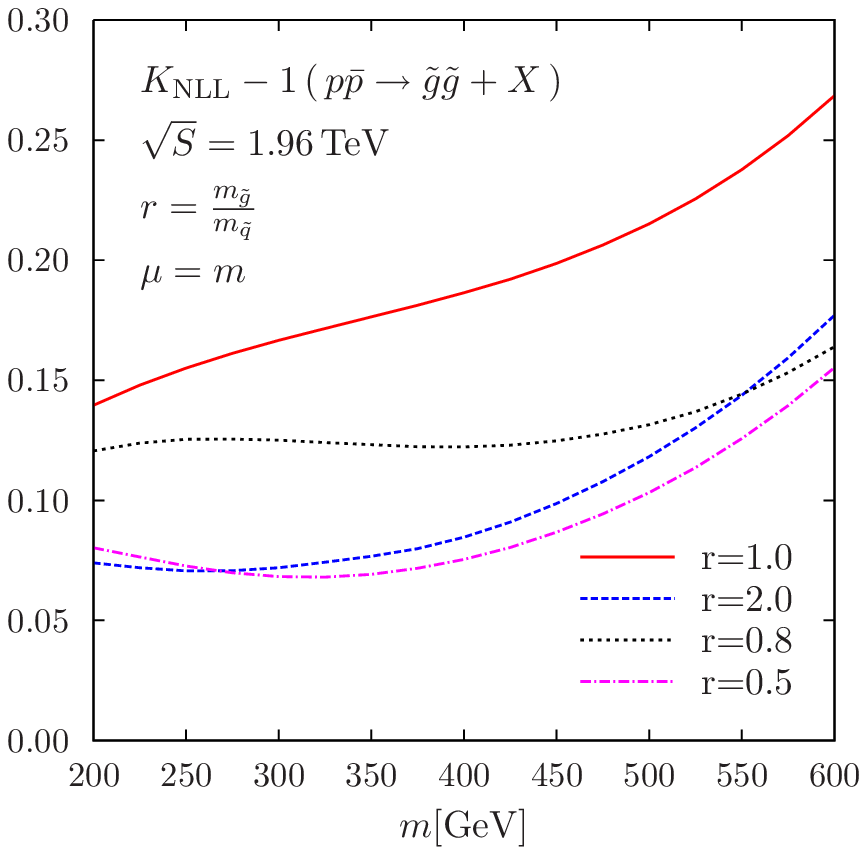, width=0.465\columnwidth }\\
(c)\epsfig{file=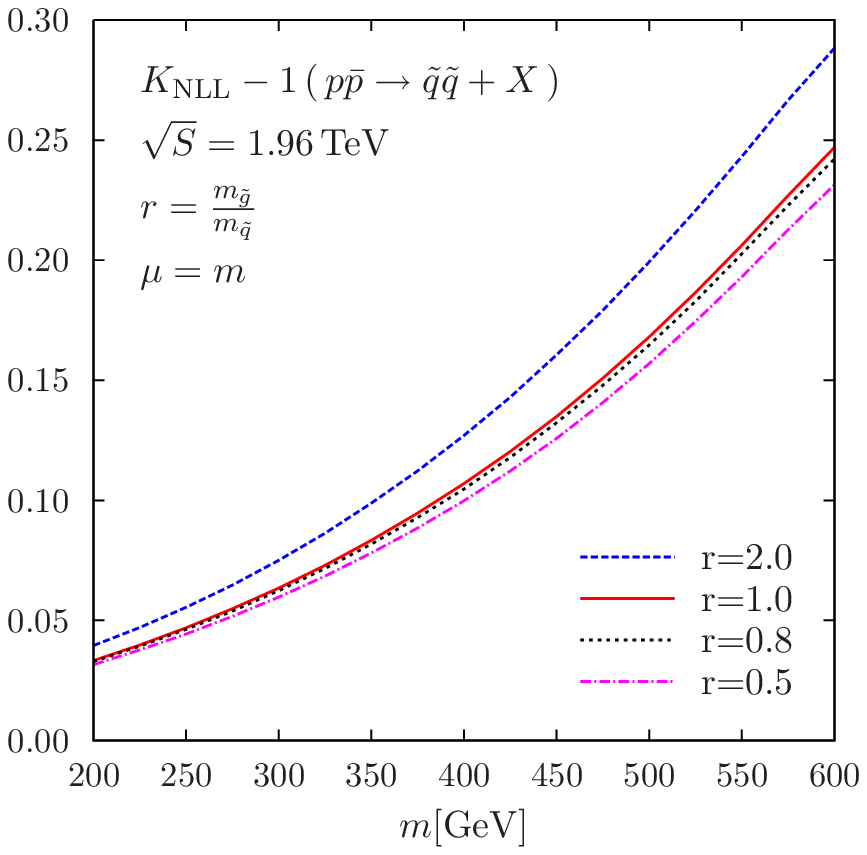, width=0.465\columnwidth}& 
(d)\epsfig{file=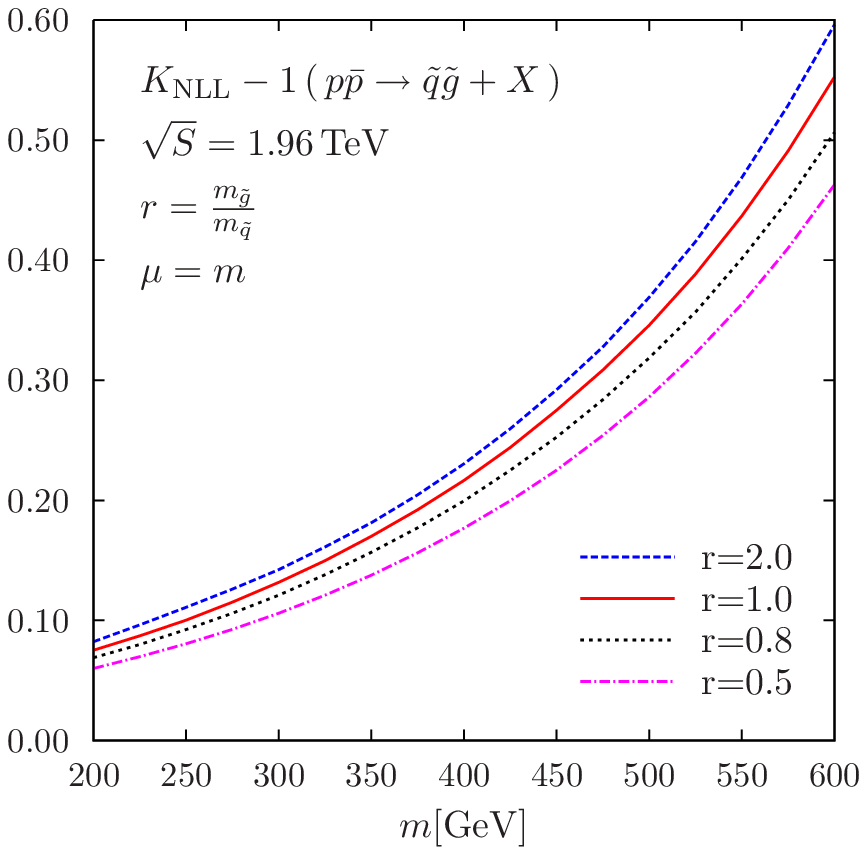, width=0.465\columnwidth }\\
  \end{tabular}
  \caption{The relative NLL $K$-factor $K_{\rm NLL} -1 = \sigma_{\rm
      NLL+NLO}/\sigma_{\rm NLO}-1$ for squark and gluino
    pair-production processes at the Tevatron as a function of the
    average sparticle mass $m$.  Shown are results for various mass
    ratios $r = m_{\tilde{g}}/m_{\tilde{q}}$.}
\label{fig:k_tev}
 }

\FIGURE{
\hspace{-1.0cm}
\begin{tabular}{ll}
(a)\epsfig{file=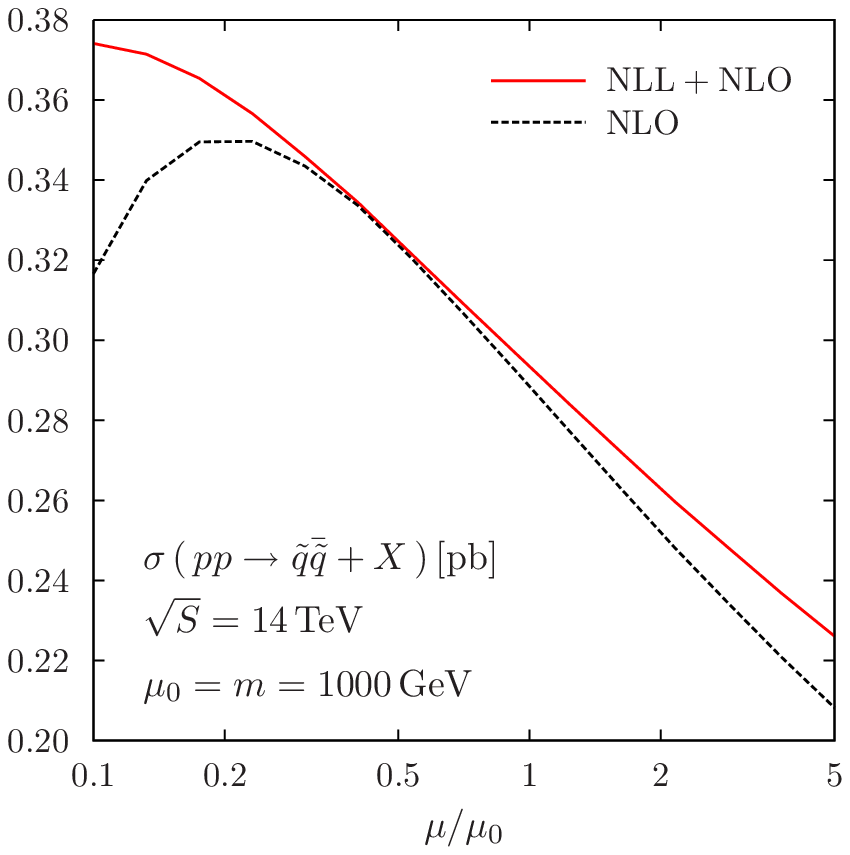, width=0.465\columnwidth}& 
(b)\epsfig{file=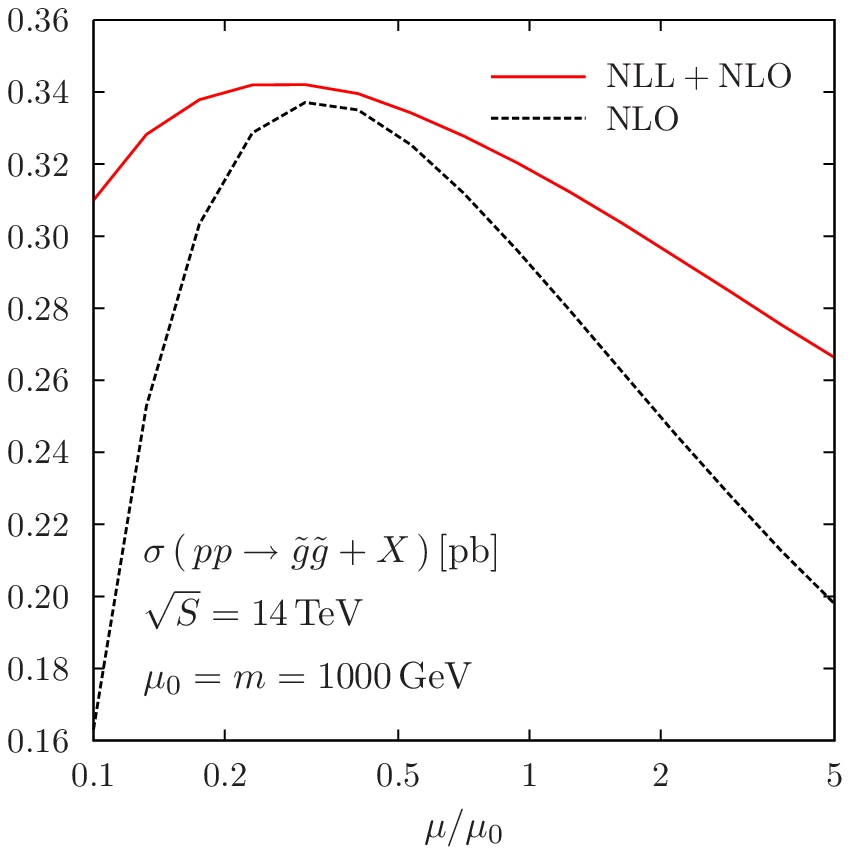, width=0.465\columnwidth }\\
(c)\epsfig{file=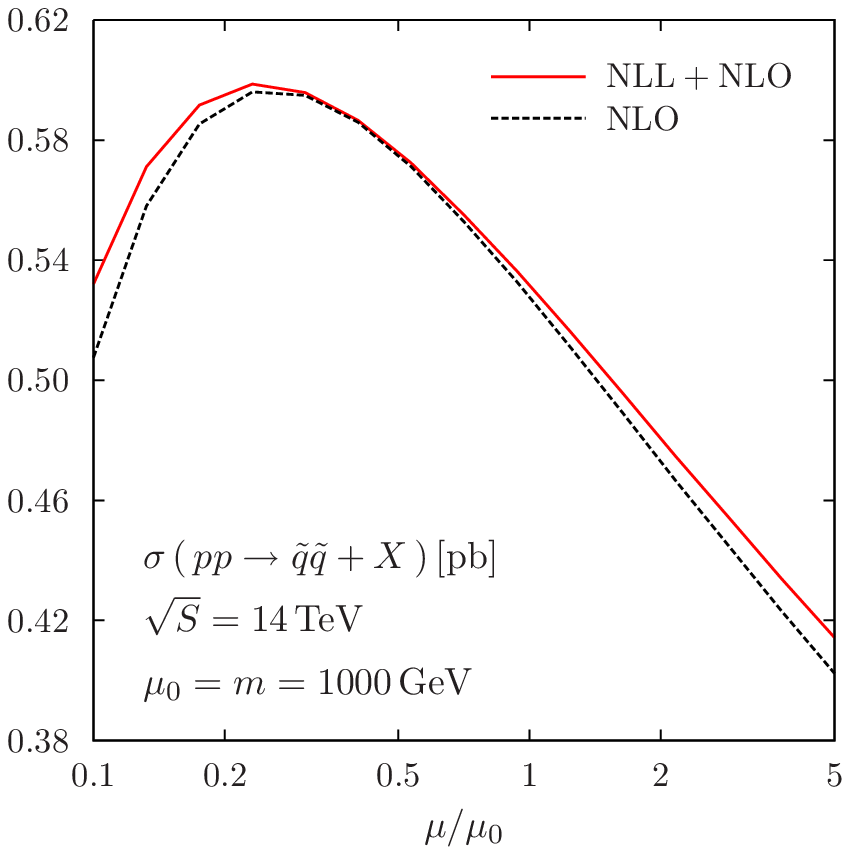, width=0.465\columnwidth}& 
(d)\epsfig{file=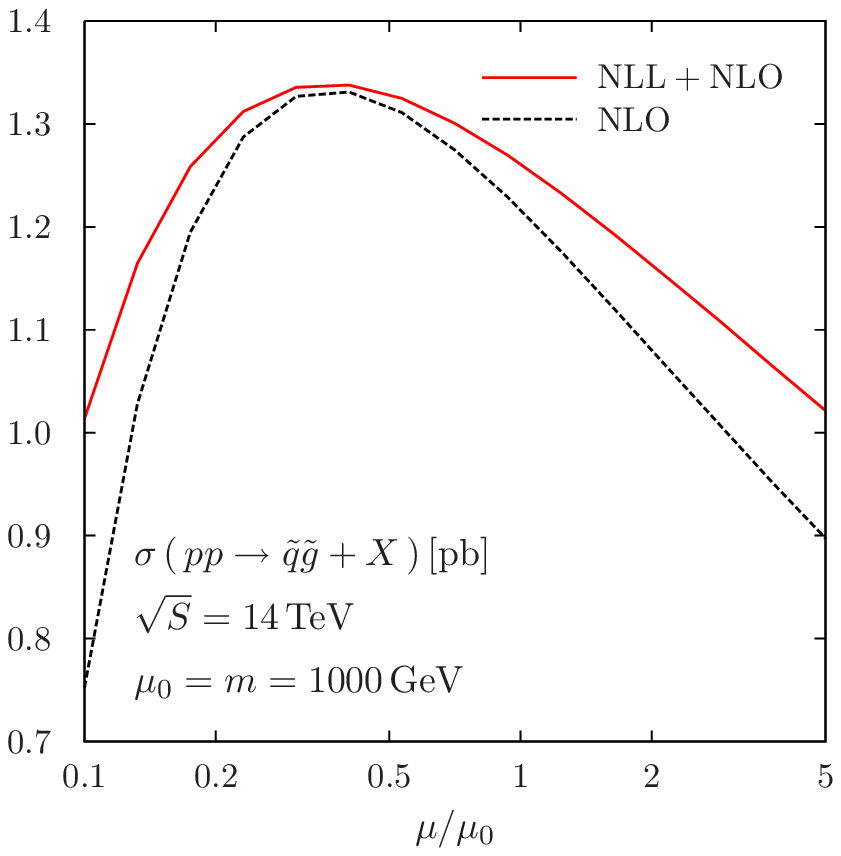, width=0.465\columnwidth }\\
\end{tabular}
\caption{The scale dependence of the NLL+NLO and the NLO total cross
  sections for squark and gluino pair-production processes at the LHC.
  The squark and gluino masses have been set to
  $m_{\sq}=m_{\gl}=m=1$~TeV.}
\label{fig:scale_lhc}
}

\FIGURE{
\hspace{-1.0cm}
\begin{tabular}{ll}
(a)\epsfig{file=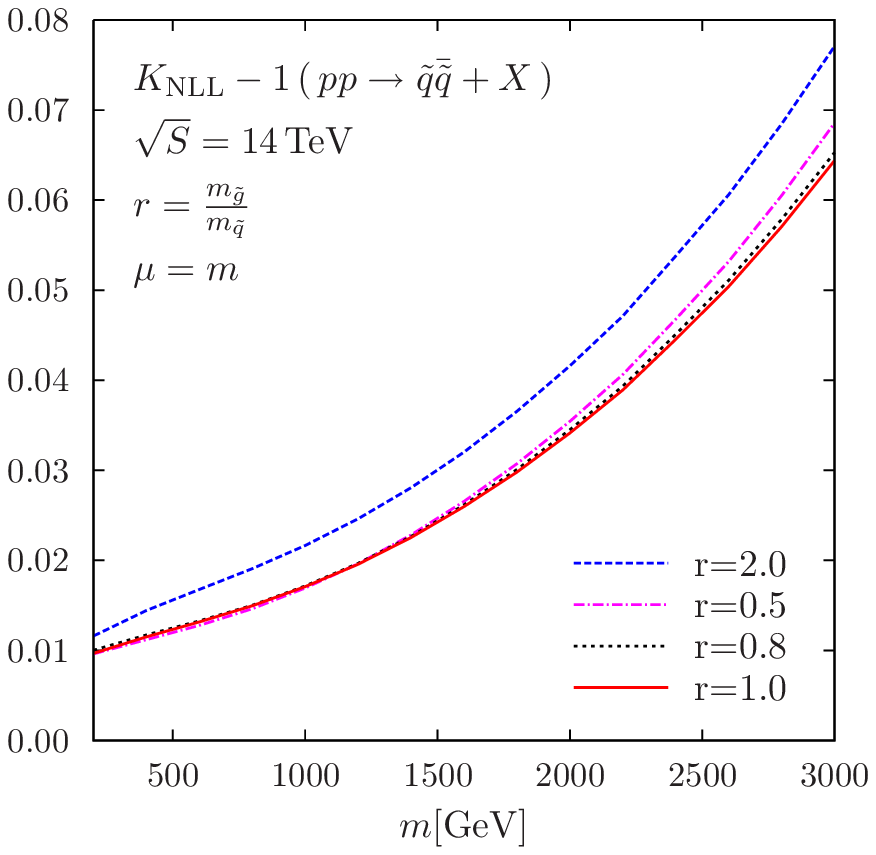, width=0.465\columnwidth} & 
(b)\epsfig{file=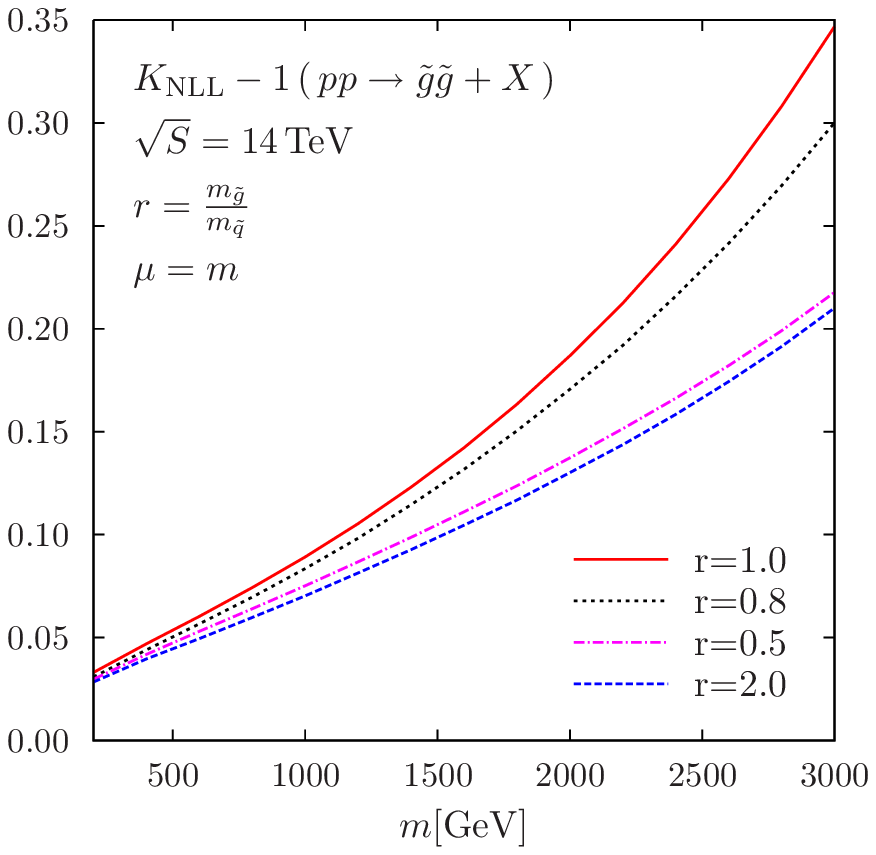, width=0.465\columnwidth }\\
(c)\epsfig{file=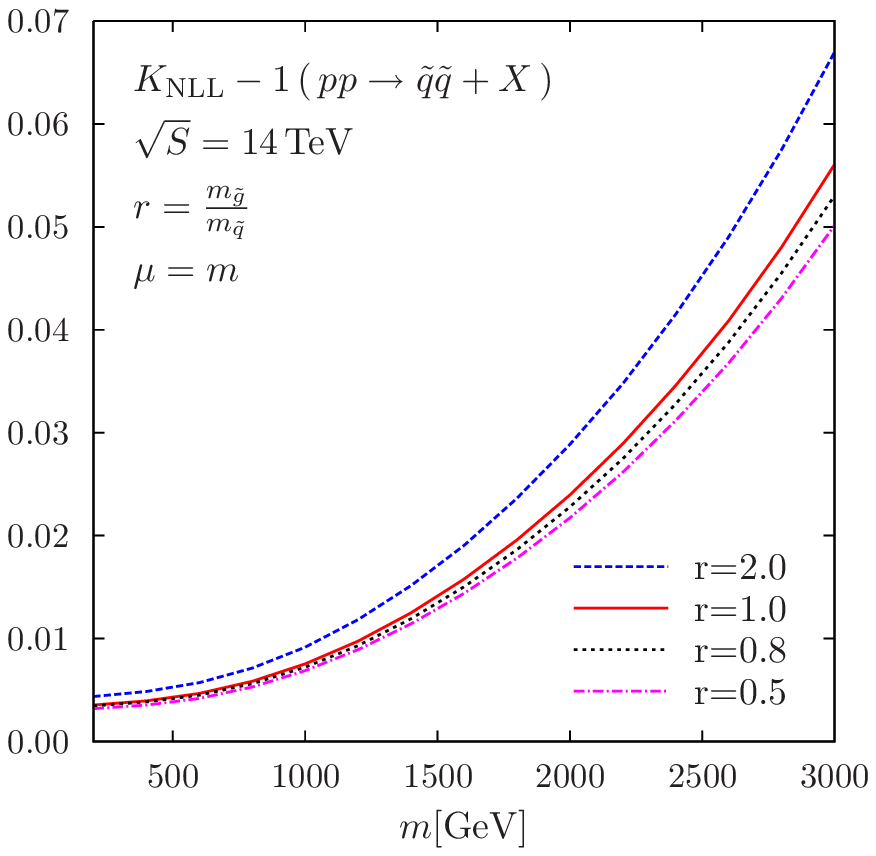, width=0.465\columnwidth}& 
(d)\epsfig{file=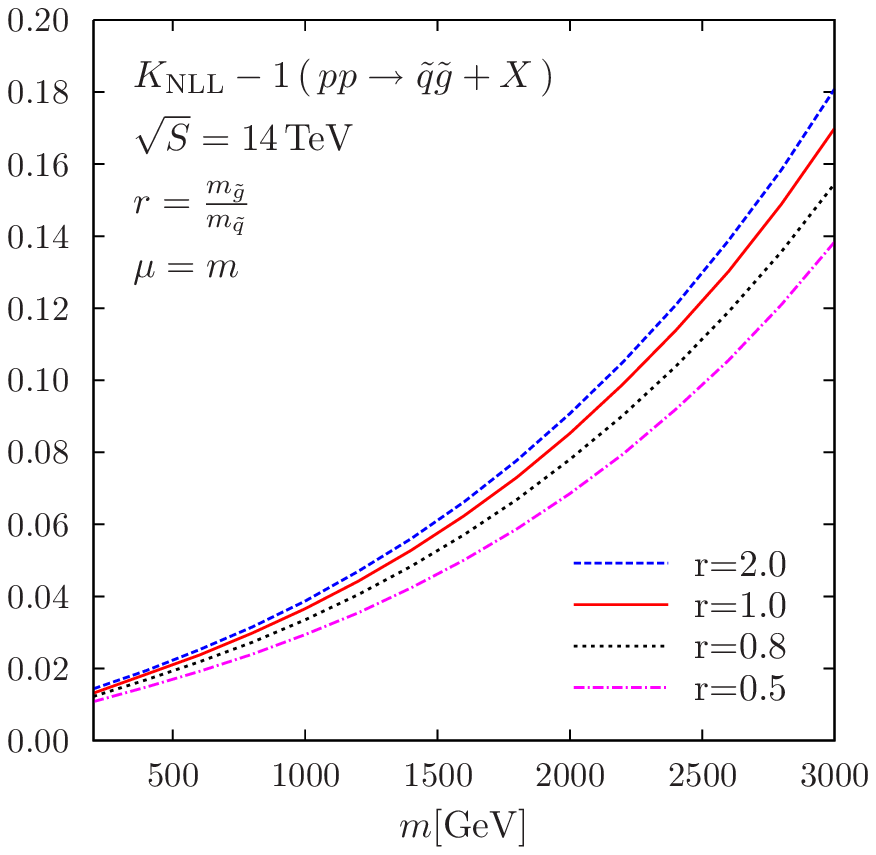, width=0.465\columnwidth }\\
\end{tabular}
\caption{The relative NLL $K$-factor $K_{\rm NLL} -1 = \sigma_{\rm
    NLL+NLO}/\sigma_{\rm NLO}-1$ for squark and gluino pair-production
  processes at the LHC as a function of the average sparticle mass
  $m$.  Shown are results for various mass ratios $r =
  m_{\tilde{g}}/m_{\tilde{q}}$.}
\label{fig:k_lhc}
 }

\FIGURE{
\hspace{-1.0cm}
\begin{tabular}{ll}
(a)\epsfig{file=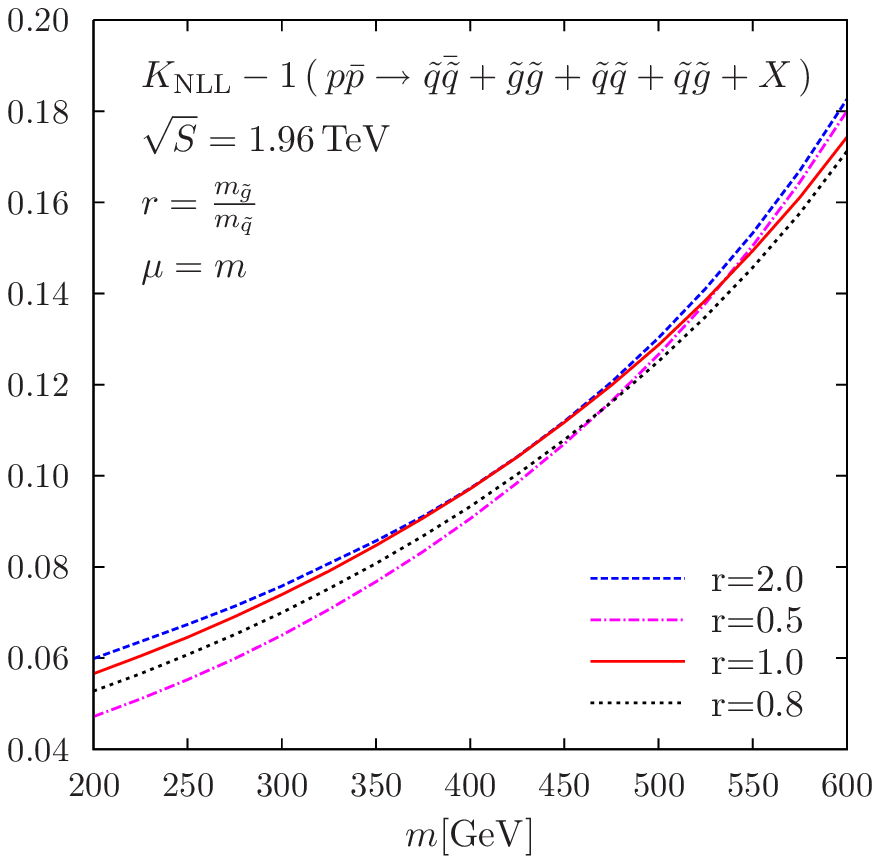, width=0.465\columnwidth} & 
(b)\epsfig{file=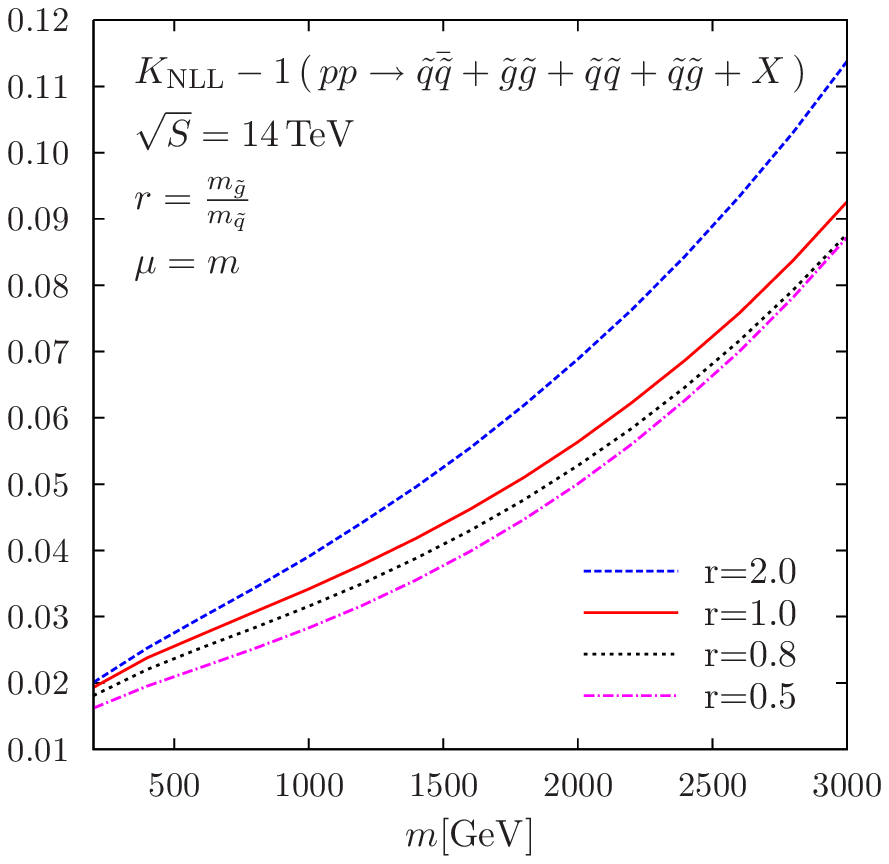, width=0.465\columnwidth}
\end{tabular}
\caption{The relative NLL $K$-factor $K_{\rm NLL} -1 =
  \sigma_{\rm NLL+NLO}/\sigma_{\rm NLO}-1$ for the inclusive squark
  and gluino pair-production cross section, $p\bar{p}/pp \to
  \tilde{q}\tilde{q}+
  \tilde{q}\bar{\tilde{q}}+\tilde{q}\tilde{g}+\tilde{g}\tilde{g} + X$,
  at the Tevatron (a) and the LHC (b) as a function of the average
  sparticle mass $m$. Shown are results for various mass ratios $r
  = m_{\tilde{g}}/m_{\tilde{q}}$.}
\label{fig:k_inclusive}
 }

\FIGURE{
\hspace{-1.0cm}
\begin{tabular}{ll}
(a)\epsfig{file=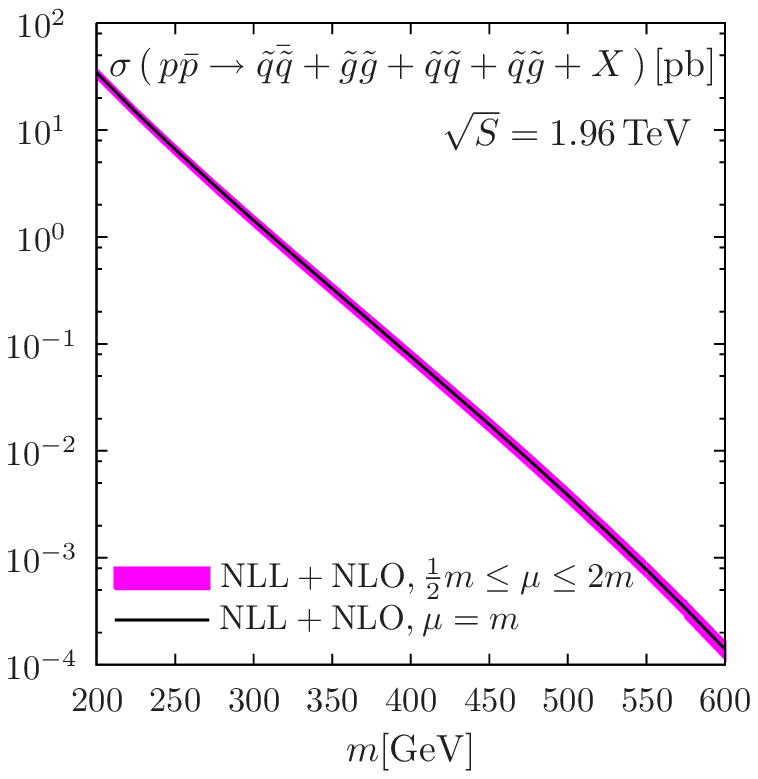, width=0.465\columnwidth}& 
(b)\epsfig{file=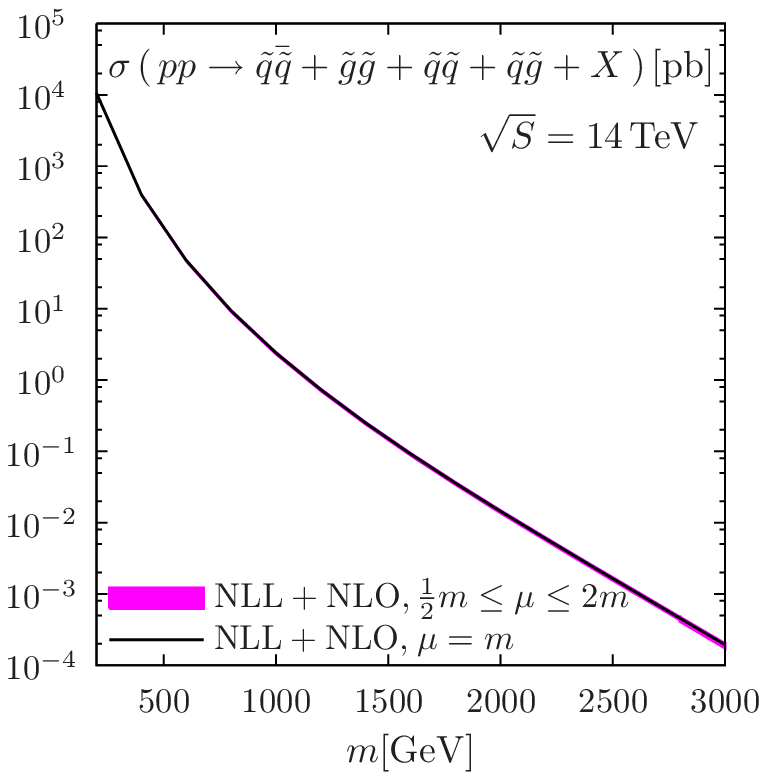, width=0.465\columnwidth}
\end{tabular}
\caption{The NLL+NLO cross section for inclusive squark and gluino
  pair-production, $p\bar{p}/pp \to \tilde{q}\tilde{q}+
  \tilde{q}\bar{\tilde{q}}+\tilde{q}\tilde{g}+\tilde{g}\tilde{g} + X$,
  at the Tevatron (a) and the LHC (b) as a function of the average
  sparticle mass $m$. Shown are results for the mass ratio $r =
  m_{\tilde{g}}/m_{\tilde{q}} =1$.  The error band corresponds to a
  variation of the common renormalization and factorization scale in
  the range $m/2\le\mu\le 2m$.}
\label{fig:total:matched}
}

\FIGURE{
\hspace{-1.0cm}
\begin{tabular}{ll}
(a)\epsfig{file=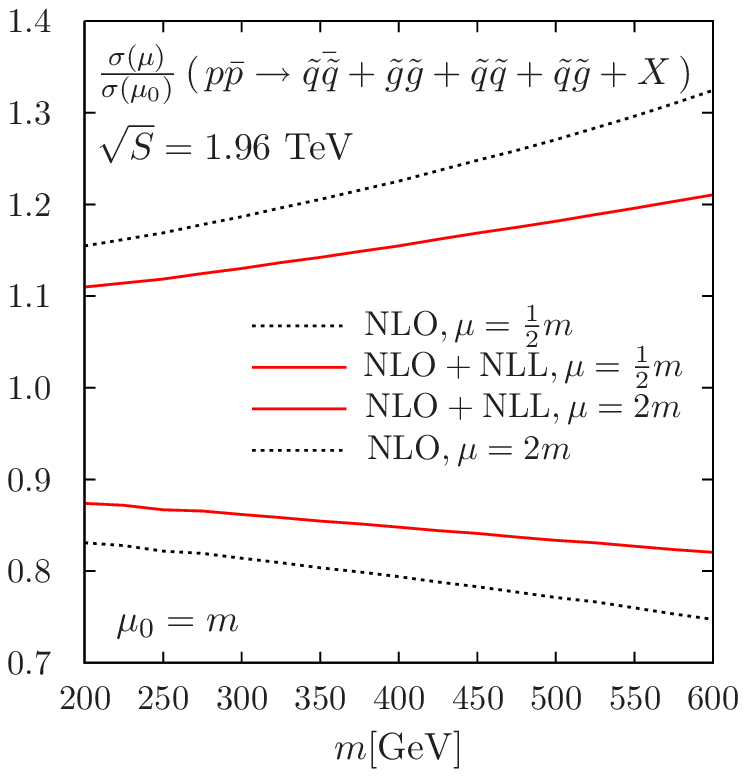, width=0.465\columnwidth}& 
(b)\epsfig{file=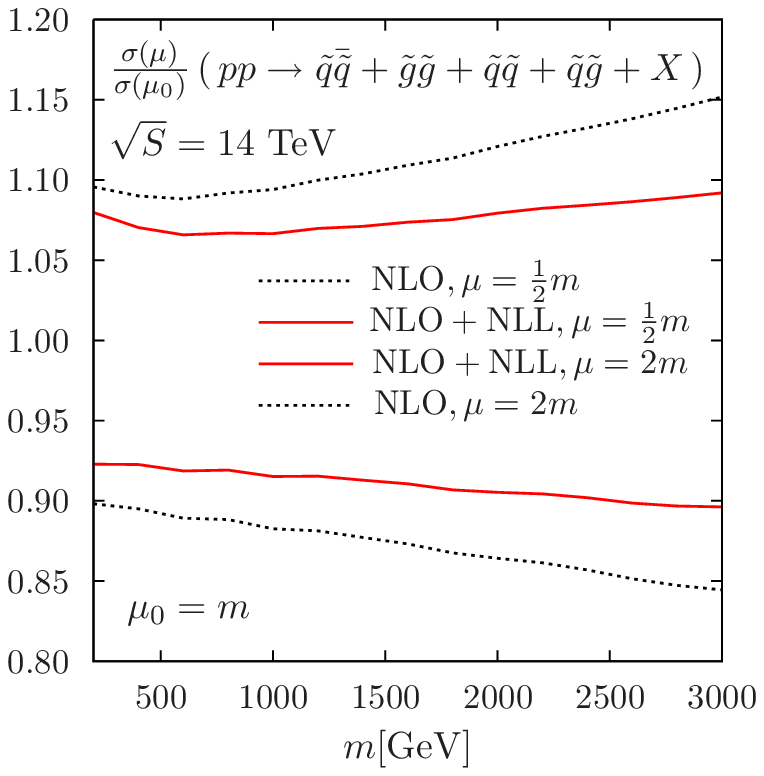, width=0.465\columnwidth}
\end{tabular}
\caption{Scale dependence of the NLL+NLO and NLO cross sections for
  inclusive squark and gluino pair-production, $p\bar{p}/pp \to
  \tilde{q}\tilde{q}+\tilde{q}\bar{\tilde{q}}
  +\tilde{q}\tilde{g}+\tilde{g}\tilde{g} + X$, at the Tevatron (a) and
  the LHC (b) as a function of the average sparticle mass $m$. Shown
  are results for the mass ratio $r = m_{\tilde{g}}/m_{\tilde{q}} =1$.
  The upper two curves correspond to the common renormalization and
  factorization scale set to $\mu = m/2$, the lower two curves to $\mu
  = 2m$.}
\label{fig:total:scale}
}

\TABLE{
\begin{small}
 \begin{tabular}{c|c|c|c|c|c}
 \multicolumn{5}{c}{\normalsize $p\bar{p} \to \tilde{q}\bar{\tilde{q}}$ at $\sqrt{S}=1.96\TeV$ (r=1.0)}
 \\[1mm]  \hline $m_{\tilde{q}} \; [\mathrm{GeV}]$ &
 200 & 300 & 400 & 500 & 600 \\  \hline \hline
$\sigma_\mathrm{NLO} \; [\mathrm{pb}]$ & $1.28\times 10^{1}$ & $7.35\times 10^{-1}$ & $4.70\times 10^{-2}$ & $2.59 \times 10^{-3}$ & $9.79\times 10^{-5}$\\  \hline
$\sigma_\mathrm{NLL+NLO} \; [\mathrm{pb}]$ & $1.30\times 10^{1}$ & $7.55\times 10^{-1}$ & $4.91\times 10^{-2}$ & $2.77\times 10^{-3}$ & $1.09\times 10^{-4}$ \\  \hline
$\mathrm{K}_{\mathrm{NLL}}-1$ & 0.016 & 0.026 & 0.045 & 0.071 & 0.11 \\
 \multicolumn{5}{c}{\normalsize \rule{0pt}{4ex} $p\bar{p} \to \tilde{g}{\tilde{g}}$ at $\sqrt{S}=1.96\TeV$ (r=1.0)}
 \\[1mm]  \hline $m_{\tilde{g}} \; [\mathrm{GeV}]$ &
 200 & 300 & 400 & 500 & 600 \\  \hline \hline
$\sigma_\mathrm{NLO} \; [\mathrm{pb}]$ &
 3.72 & $1.07\times 10^{-1}$ & $4.61\times 10^{-3}$ & $1.96\times 10^{-4}$ & $6.01\times 10^{-6}$\\  \hline
$\sigma_\mathrm{NLL+NLO} \; [\mathrm{pb}]$ &
 4.24 & $1.24\times 10^{-1}$ & $5.47\times 10^{-3}$ & $2.38\times 10^{-4}$ & $7.62\times 10^{-6}$ \\  \hline
$\mathrm{K}_{\mathrm{NLL}}-1$ &
 0.14 & 0.17 & 0.19 & 0.22 & 0.27 \\
 \multicolumn{5}{c}{\normalsize \rule{0pt}{4ex} $p\bar{p} \to \tilde{q}\tilde{q}$ at $\sqrt{S}=1.96\TeV$ (r=1.0)}
 \\[1mm]  \hline $m_{\tilde{q}} \; [\mathrm{GeV}]$ &
 200 & 300 & 400 & 500 & 600 \\  \hline \hline
$\sigma_\mathrm{NLO} \; [\mathrm{pb}]$ &
 1.81 &  $4.78\times 10^{-2}$ & $1.39\times 10^{-3}$ & $3.38\times 10^{-5}$ & $5.66\times 10^{-7}$\\  \hline
$\sigma_\mathrm{NLL+NLO} \; [\mathrm{pb}]$ &
 1.87 & $5.09\times 10^{-2}$ & $1.54\times 10^{-3}$ & $3.95\times 10^{-5}$ & $7.06\times 10^{-7}$ \\  \hline
$\mathrm{K}_{\mathrm{NLL}}-1$ &
 0.033 & 0.064 & 0.11 & 0.17 & 0.25 \\
 \multicolumn{5}{c}{\normalsize \rule{0pt}{4ex} $p\bar{p} \to \tilde{q}{\tilde{g}}$ at $\sqrt{S}=1.96\TeV$ (r=1.0)}
 \\[1mm]  \hline $m \; [\mathrm{GeV}]$ &
 200 & 300 & 400 & 500 & 600 \\  \hline \hline 
$\sigma_\mathrm{NLO} \;[\mathrm{pb}]$ &
 $1.43\times 10^{1}$ & $4.44\times 10^{-1}$ & $1.71\times 10^{-2}$ & $5.98\times 10^{-4}$ & $1.46\times 10^{-5}$ \\  \hline
$\sigma_\mathrm{NLL+NLO} \; [\mathrm{pb}]$ &
 $1.54\times 10^{1}$ & $5.03\times 10^{-1}$ & $2.09\times 10^{-2}$ & $8.05\times 10^{-4}$ & $2.27\times 10^{-5}$ \\  \hline
$\mathrm{K}_{\mathrm{NLL}}-1$ &
 0.075 & 0.13 & 0.22 & 0.35 & 0.55
 \end{tabular}

\vspace*{2mm}
\caption{The NLL+NLO and NLO cross sections for the squark and gluino
  pair-production processes at the Tevatron.  Shown are results for
  the mass ratio $r = m_{\tilde{g}}/m_{\tilde{q}} =1$. The 
  common renormalization and factorization scale has been set to $m$.}
\label{tab:tev}
\end{small}
}

\TABLE{
\begin{small}
 \begin{tabular}{c|c|c|c|c|c}
 \multicolumn{5}{c}{\normalsize $pp \to \tilde{q}\bar{\tilde{q}}$ at $\sqrt{S}=14\TeV$ (r=1.0)}
 \\[1mm]  \hline $m_{\tilde{q}} \; [\mathrm{GeV}]$ &
 200 & 500 & 1000 & 2000 & 3000 \\  \hline \hline
$\sigma_\mathrm{NLO}\; [\mathrm{pb}]$ &
 $1.30\times 10^{3}$ & $1.60\times 10^{1}$ & $2.89\times 10^{-1}$ & $1.11\times 10^{-3}$ & $7.13\times 10^{-6}$  \\  \hline
$\sigma_\mathrm{NLL+NLO} \; [\mathrm{pb}]$ & 
 $1.31\times 10^{3}$ & $1.61\times 10^{1}$ & $2.93\times 10^{-1}$ & $1.14\times 10^{-3}$ & $7.59\times 10^{-6}$  \\  \hline
$\mathrm{K}_{\mathrm{NLL}}-1$ & 
 0.010 & 0.012 & 0.017 &  0.034 & 0.064  \\
 \multicolumn{5}{c}{\normalsize \rule{0pt}{4ex} $pp \to \tilde{g}\tilde{g}$ at $\sqrt{S}=14\TeV$ (r=1.0)}
 \\[1mm]  \hline $m_{\tilde{g}} \; [\mathrm{GeV}]$ &
 200 & 500 & 1000 & 2000 & 3000 \\  \hline \hline
$\sigma_\mathrm{NLO} \; [\mathrm{pb}]$ &
 $3.74\times 10^{3}$ & $2.85\times 10^{1}$ & $2.92\times 10^{-1}$ & $5.82\times 10^{-4}$ & $2.68\times 10^{-6}$  \\  \hline
$\sigma_\mathrm{NLL+NLO} \; [\mathrm{pb}]$ & 
 $3.86\times 10^{3}$ & $3.00\times 10^{1}$ & $3.18\times 10^{-1}$ & $6.91\times 10^{-4}$ & $3.62\times 10^{-6}$  \\  \hline
$\mathrm{K}_{\mathrm{NLL}}-1$ & 
 0.033 & 0.054 & 0.089 & 0.19 & 0.35  \\
 \multicolumn{5}{c}{\normalsize \rule{0pt}{4ex} $pp \to \tilde{q}\tilde{q}$ at $\sqrt{S}=14\TeV$ (r=1.0)}
 \\[1mm]  \hline $m_{\tilde{q}} \; [\mathrm{GeV}]$ &
 200 & 500 & 1000 & 2000 & 3000 \\  \hline \hline
$\sigma_\mathrm{NLO} \; [\mathrm{pb}]$ &
 $5.45\times 10^{2}$ & $1.34\times 10^{1}$ & $5.28\times 10^{-1}$ & $6.48\times 10^{-3}$ & $1.18\times 10^{-4}$  \\  \hline
$\sigma_\mathrm{NLL+NLO} \; [\mathrm{pb}]$ & 
 $5.46\times 10^{2}$ & $1.34\times 10^{1}$ & $5.32\times 10^{-1}$ & $6.64\times 10^{-3}$ & $1.25\times 10^{-4}$  \\  \hline
$\mathrm{K}_{\mathrm{NLL}}-1$ & 
 0.003 &  0.004 & 0.008 & 0.024 & 0.056  \\
 \multicolumn{5}{c}{\normalsize \rule{0pt}{4ex} $pp \to \tilde{q}\tilde{g}$ at $\sqrt{S}=14\TeV$ (r=1.0)}
 \\[1mm]  \hline $m \; [\mathrm{GeV}]$ &
 200 & 500 & 1000 & 2000 & 3000 \\  \hline \hline
$\sigma_\mathrm{NLO} \; [\mathrm{pb}]$ &
 $4.86\times 10^{3}$ & $6.55\times 10^{1}$ & $1.22$  & $5.49\times 10^{-3}$ & $4.96\times 10^{-5}$  \\  \hline
$\sigma_\mathrm{NLL+NLO} \; [\mathrm{pb}]$ & 
 $4.92\times 10^{3}$ & $6.69\times 10^{1}$ & $1.26$ & $5.96\times 10^{-3}$ & $5.80\times 10^{-5}$  \\  \hline
$\mathrm{K}_{\mathrm{NLL}}-1$ & 
 0.013 & 0.021 & 0.037 & 0.085 & 0.17  \\
 \end{tabular}

\vspace*{2mm}
\caption{The NLL+NLO and NLO cross sections for the squark and gluino
  pair-production processes at the LHC.  Shown are results for the
  mass ratio $r = m_{\tilde{g}}/m_{\tilde{q}} =1$. The common
  renormalization and factorization scale has been set to $m$.}
\label{tab:lhc}
\end{small}
}

\clearpage

\begin{appendix}

\section{\boldmath Leading-order $N$-space cross sections for $\sq\sq$ and 
         $\sq\gl$ production}
\label{sec:appA}

In this appendix we present the analytical results for the Mellin
transforms of the LO cross sections for $\sq\sq$ and $\sq\gl$
production. The cross sections are colour-decomposed in SU(3)
according to the procedure described in
section~\ref{se:anomalous_dim}. The Mellin-transformed LO cross
sections for the $\sq\sqb$ and $\gl\gl$ final states can be found
in~\cite{Kulesza:2009kq}.

The expressions for the colour-decomposed LO $N$-space cross sections for the 
process $q_{f_1} q_{f_2} \to \sq\sq\,$ are given by
\begin{eqnarray}
\tilde\si^{(0)}_{qq \to \sq\sq,1}(N) 
&=& \frac{\alpha_{\rm s}^2\pi}{27 m_{\sq}^2}
    \Bigg[ -\, \delta_{f_1f_2} H_N 
           \,-\, \frac{4B_NG_N}{2N+3}\,\Bigl(N+\frac{2r^2}{r^2+1}\,
                 \frac{1}{N+2}\,\Bigr) \nn \\[3mm]
&& \qquad\qquad\qquad
   {}+\, 2B_N\,\frac{N^2+2N+2}{(N+1)(N+2)}\, \Bigg]\,, \\[3mm]
\tilde\si^{(0)}_{qq \to \sq\sq,2}(N) 
&=& \frac{\alpha_{\rm s}^2\pi}{27 m_{\sq}^2}
    \Bigg[ \delta_{f_1f_2} \frac{H_N}{2} 
           \,-\, \frac{2B_NG_N}{2N+3}\,\Bigl(N+\frac{2r^2}{r^2+1}\,
                 \frac{1}{N+2}\,\Bigr) \nn \\[3mm]
&& \qquad\qquad\qquad
   {}+\, B_N\,\frac{N^2+2N+2}{(N+1)(N+2)}\, \Bigg]\,,
\end{eqnarray}
whereas for the process $\,qg \to \sq\gl\,$ they read
\begin{eqnarray}
\tilde\si^{(0)}_{qg \to \sq\gl,1}(N) 
&=& \frac{\alpha_{\rm s}^2\pi }{8 m_{\sq}^2} 
    \left[ \frac{9\,B_{N+1}P^-_{N+1}\,(1-r)}{(r+1)^3} 
           \,-\, \frac{9\,B_{N}P^-_{N}}{2(r+1)^2} 
           \,+\, \frac{B_{N+2}P^-_{N+2}\,(7r^2-9)(1-r)}{(r+1)^5} \right. 
    \nn \\[3mm]
&+& \left. \frac{ B_{N+1}P^+_{N+1}\,(1-r)}{9(r+1)^3}
           \,-\, \frac{B_{N+2}P^+_{N+2}\,(r^2+17)(1-r)}{9(r+1)^5} \right. 
    \nn \\[3mm]
&+& \left. \frac{130\,B_{N+1}K_{N+1}\,(1-r)}{9(r+1)^3} 
           \,-\, \frac{56\,B_N K_{N}}{9(r+1)^2}
    \right]\,, \\[3mm]
\tilde\si^{(0)}_{qg \to \sq\gl,2}(N) 
&=& \frac{\alpha_{\rm s}^2\pi}{8 m_{\sq}^2} 
    \left[ \frac{2\,B_{N+1}P^-_{N+1}\,(1-r)}{(r+1)^3} 
           \,-\,\frac{B_N P^-_{N}}{(r+1)^2}
           \,-\, \frac{2\,B_{N+2}P^-_{N+2}\,(r^2+1)(1-r)}{(r+1)^5}\right. 
    \nn \\[3mm]
&+& \left. \frac{2\,B_{N+1}P^+_{N+1}\,(1-r)}{(r+1)^3}
           \,-\, \frac{2\,B_{N+2}P^+_{N+2}\,(r^2+1)(1-r)}{(r+1)^5} \right. 
    \nn \\[3mm]
&+& \left. \frac{4\,B_{N+1}K_{N+1}\,(1-r)}{(r+1)^3}
    \right]\,, \\[3mm]
\tilde\si^{(0)}_{qg \to \sq\gl,3}(N) 
&=& \frac{5}{2}\,\tilde\si^{(0)}_{qg \to \sq\gl,2}(N)\,.
\end{eqnarray}

We have used the following abbreviations:
\begin{eqnarray}
B_{N} &\equiv& \beta (N+1, 1/2) \,, \nn \\[3mm]
G_{N} &\equiv& _2 F_1 \left( 1,1/2,N+5/2,
  \left(\frac{r^2-1}{r^2+1}\right)^2 \right) \,, 
\nn \\
P_N^{\pm} &\equiv&
\frac{-1}{N+1} \  _2 F_1 \left( 1/2, N+1, N+3/2, \left( \frac{1-r}{1+r}
  \right)^2 \right)  
\nn \\
&\pm&  \left(\frac{1-r}{r+1}\right) \frac{1}{N+3/2} \
_2 F_1 \left( 1/2, N+2, N+5/2, \left( \frac{1-r}{1+r}  \right)^2
\right) \,, 
\nn \\[2mm]
K_N &\equiv& \frac{1}{2N+3} \
_2 F_1 \left( -1/2, N+1, N+5/2, \left( \frac{1-r}{1+r}  \right)^2 \right) \,,
\nn \\[2mm]
H_N &\equiv& \int_0^1 d z \frac{z^{N+1}}{\frac{1}{r^2}- 
             \left(\frac{1-r^2}{2r^2}\right)z}
             \log\left( \frac{2(1+\sqrt{1-z})+(r^2-1)z}
                             {2(1-\sqrt{1-z})+(r^2-1)z} \right) \,, \nn \\
\end{eqnarray}
with $_2F_1(\lambda,\mu,\nu,\xi)$ the hypergeometric function,
$\beta(\mu,\nu)$ the beta function and $r=m_{\tilde g}/m_{\tilde q}$.
For the numerical evaluation of $H_N$ we use the expansion
\begin{eqnarray}
H_N  &=&  \frac{2r^2}{1+r^2}\sum_{m=0}^{\infty}
          \left( \frac{ r^2 -1}{1+r^2}\right)^m
          \frac{1}{1+m}\,\sum_{k=0}^{m} \,
          \frac{\left(-1\right)^k}{\beta\left(k+1,m-k+1\right)}
\nn  \\[3mm]
&\times& \left[ \frac{\beta \left(k+N+2, 1/2 \right)}{k+N+2} -2
                \left(\frac{r^2 -1}{1+r^2}\right)\beta \left(k+N+2,3/2\right) 
         \right.
\nn \\[3mm]
&& \left. _2F_1 \left(1,1/2,k+N+7/2,\left( \frac{1-r^2}{1+r^2}\right)^2\right)
   \right]\,.
\end{eqnarray}

\section{\boldmath Construction of the $s$-channel colour basis: an
  example}
\label{sec:appB} 

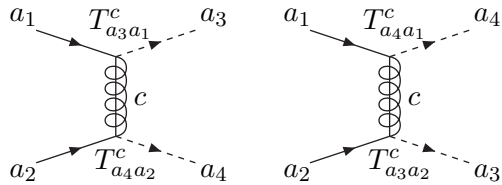
\begin{figure}[!h]
\begin{center}
  \begin{picture}(100,80)
    \SetWidth{0.5}
    \Gluon(40,50)(40,20){4}{4}
    \Line(40,50)(40,20)
    \ArrowLine(10,60)(40,50)
    \ArrowLine(10,10)(40,20)
    \DashArrowLine(40,20)(70,10){2}
    \DashArrowLine(40,50)(70,60){2}
    \Text(0,3)[lb]{$a_2$}
    \Text(72,62)[lb]{$a_3$}
    \Text(72,3)[lb]{$a_4$}
    \Text(0,62)[lb]{$a_1$}
    \Text(30,57)[lb]{$T^c_{a_3a_1}$}
    \Text(32,4)[lb]{$T^c_{a_4a_2}$}
    \Text(47,32)[lb]{$c$}
  \end{picture}
  \begin{picture}(100,80)
    \SetWidth{0.5}
    \Gluon(40,50)(40,20){4}{4}
    \Line(40,50)(40,20)
    \ArrowLine(10,60)(40,50)
    \ArrowLine(10,10)(40,20)
    \DashArrowLine(40,20)(70,10){2}
    \DashArrowLine(40,50)(70,60){2}
    \Text(0,3)[lb]{$a_2$}
    \Text(72,62)[lb]{$a_4$}
    \Text(72,3)[lb]{$a_3$}
    \Text(0,62)[lb]{$a_1$}
    \Text(30,57)[lb]{$T^c_{a_4a_1}$}
    \Text(32,4)[lb]{$T^c_{a_3a_2}$}
    \Text(47,32)[lb]{$c$}
  \end{picture}
  \caption{The LO diagrams that contribute to squark-pair production.
           \label{f:qqtoqqprocess}}
  \end{center}
\end{figure}
For the process $qq\to\tilde q\tilde q$ we explicitly show how to
derive the $s$-channel colour basis given in
Eq.~(\ref{eq:qqtoqqbasis}).  The same steps can be used to obtain the
basis given in Eq.~(\ref{eq:qgtoqgbasis}) for the $qg\to\tilde q\tilde
g$ process, although the calculations are more tedious in that case.

As a starting point we take the colour structures that occur in the LO
$qq\to \tilde q\tilde q$ process displayed in
Fig.~\ref{f:qqtoqqprocess}.  Using the conventions introduced in
section~\ref{s:kinematics} these are:
\begin{align*}
T^c_{a_3a_1}T^c_{a_4a_2}&\;=\; \frac{1}{2}\left(\delta_{a_3a_2}\delta_{a_4a_1}
-\frac{1}{N_{_C}}\delta_{a_3a_1}\delta_{a_4a_2}\right)\,,\\
T^c_{a_3a_2}T^c_{a_4a_1}&\;=\;\frac{1}{2}\left(\delta_{a_3a_1}\delta_{a_4a_2}
-\frac{1}{N_{_C}}\delta_{a_3a_2}\delta_{a_4a_2}\right)\,,
\end{align*}
where $c$ is a summation index in the adjoint representation. For
convenience these colour structures have been rewritten in terms of
the $t$ and $u$-channel singlet structures
$\delta_{a_3a_1}\delta_{a_4a_2}$ and $\delta_{a_3a_2}\delta_{a_4a_1}$.
It is clear from this expression that two independent singlet
structures occur. Since the two-particle reducible product
representation $\mathbf{3}\otimes\mathbf{3}$ contains two irreducible
representations, cf.\ Eq.~(\ref{eq:irreps}), this basis must be
complete.  That means that the $s$-channel base tensors are linear
combinations of these singlet structures. The projective prescription
(\ref{eq:selfprojective}) leads to the following set of equations:
\[(A_I\delta_{ba_2}\delta_{b'a_1} + B_I\delta_{ba_1}\delta_{b'a_2})
(A_{I'}\delta_{a_3b'}\delta_{a_4b} +
B_{I'}\delta_{a_3b}\delta_{a_4b'}) =
Z\delta_{II'}(A_I\delta_{a_3a_2}\delta_{a_4a_1} +
B_I\delta_{a_3a_1}\delta_{a_4a_2})\,,\] where $I,I'\in\{1,2\}$ and $Z$
is an arbitrary normalization constant. Working out the equations
shows that up to interchanging the base tensors the unique solution is
given by $A_1=-A_2=B_1=B_2=Z/2$, which is exactly the basis given in
Eq.~(\ref{eq:qqtoqqbasis}).  \FIGURE{
 \begin{picture}(130,75) (-20,-5)
    \SetWidth{0.5}
    \COval(45,30)(15,15)(0){Black}{White}
    \Text(40,25)[lb]{\Large{$c_{_I}$}}
    \DashArrowLine(60,22)(90,0){2}
    \DashArrowLine(60,37)(90,60){2}
    \ArrowLine(0,60)(31,36)
    \ArrowLine(0,0)(31,23)
    \GlueArc(45,30)(30,150,32){-4}{14}
        \Text(0,11)[lb]{$a_2$}
    \Text(80,45)[lb]{$a_3$}
    \Text(80,11)[lb]{$a_4$}
    \Text(0,45)[lb]{$a_1$}
  \end{picture}
   \caption{An example of gluon insertion.\label{f:gluoninsertion}}
}

One can check explicitly that this basis is complete for gluon
resummation: representing the combined colour structure of the
external particles by one of the base tensors $c_I$ and connecting any
two external particles by an additional gluon yields no additional
colour structures.  In Fig.~\ref{f:gluoninsertion} an example of such
a gluon insertion is shown.  For processes for which the LO colour
basis is not complete, this procedure can also be used to identify
additional base tensors.

\FIGURE{
  \begin{picture}(140,60) (142,-115)
    \SetWidth{0.5}
    \ArrowLine(165,-59)(195,-89)
    \Gluon(165,-119)(195,-89){4}{5}
    \ArrowLine(195,-89)(240,-89)
    \DashArrowLine(240,-89)(270,-59){2}
    \Gluon(240,-89)(270,-119){4}{5}
    \Line(240,-89)(270,-119)
    \Text(160,-75)[lb]{$a_1$}
    \Text(160,-104)[lb]{$a_2$}
    \Text(265,-75)[lb]{$a_3$}
    \Text(265,-104)[lb]{$a_4$}
    \Text(215,-100)[lb]{$c$}
    \Text(190,-85)[lb]{$T^{a_2}_{ca_1}$}
    \Text(223,-85)[lb]{$T^{a_4}_{a_3c}$}
  \end{picture}
  \caption{Example of a diagram corresponding to a base tensor.
           \label{f:qgtoqgfundamental}}
}

If a particle is exchanged in the $s$-channel, the corresponding base
tensor has a direct physical interpretation. An example is the Feynman
diagram for the $qg\to \tilde q\tilde g$ process shown in
Fig.~\ref{f:qgtoqgfundamental}. Since the quark exchanged in the
$s$-channel is in the fundamental representation, the corresponding
$N_{_C}$-dimensional base tensor ($c_1^{qg}$ in
Eq.~(\ref{eq:qgtoqgbasis})) can be read off immediately from the
colour structure of this diagram.

\section{Eikonal Feynman rules}
\label{sec:appC}

In this appendix the eikonal Feynman rules will be given for a soft
gluon with momentum $k$ attached to an eikonal line with momentum $p$.
In the eikonal approximation we have $k\ll p$, which leads to simple
Feynman rules since the propagator that connects the matrix element to
the radiated gluon becomes effectively on-shell. The generic diagrams
and their corresponding Feynman rules are given by
(cf.~\cite{Kidonakis:1997gm})
\begin{equation}
      \parbox{63pt}{\begin{picture}(61,59)(0,-8)
    \SetWidth{0.5}
    \GOval(54,23)(7,7)(0){0.882}
    \Line(0,24)(47,24)
    \LongArrow(0,30)(20,30)
    \Text(9,35)[lb]{\large{$p$}}
    \Gluon(22,24)(47,0){3}{5}
\LongArrow(22,15)(34,4)
    \Text(22,2)[lb]{\large{$k$}}
    \Text(1,17)[lb]{$a$}
    \Text(41,27)[lb]{$b$}
     \Text(44,-8)[lb]{$\mu,c$}
  \end{picture}}\ =\ g_s(T^c_R)_{ab}\frac{p_\mu}{p\cdot k- i\epsilon} 
      \parbox{55pt}{\begin{picture}(50,59)(12,-8)
    \SetWidth{0.5}
    \GOval(54,23)(7,7)(0){0.882}
    \Line(22,24)(47,24)
    \LongArrow(22,30)(42,30)
    \Text(31,35)[lb]{\large{$p$}}
    \Text(41,15)[lb]{$b$}
  \end{picture}}\label{eq:ineikonaldiagram}\vspace*{2mm}
\end{equation}
for an incoming eikonal line and
\begin{equation}
      \parbox{63pt}{\begin{picture}(61,59)(0,-8)
    \SetWidth{0.5}
    \GOval(0,23)(7,7)(0){0.882}
    \Line(7,24)(54,24)
    \LongArrow(33,30)(53,30)
    \Text(43,35)[lb]{\large{$p$}}
    \Gluon(26,24)(51,0){3}{5}
\LongArrow(26,15)(38,4)
    \Text(26,2)[lb]{\large{$k$}}
    \Text(8,27)[lb]{$b$}
    \Text(48,17)[lb]{$a$}
    \Text(46,-8)[lb]{$\mu,c$}
  \end{picture}}\ =\ g_s(T^c_R)_{ab}\frac{p_\mu}{p\cdot k+ i\epsilon} 
      \parbox{55pt}{\begin{picture}(50,59)(12,-8)
    \SetWidth{0.5}
    \GOval(26,23)(7,7)(0){0.882}
    \Line(33,24)(58,24)
    \LongArrow(33,30)(53,30)
    \Text(38,35)[lb]{\large{$p$}}
    \Text(35,15)[lb]{$b$}
  \end{picture}}\label{eq:outeikonaldiagram}\vspace*{2mm}
\end{equation}
for an outgoing eikonal line.  Here $g_s$ is the strong coupling
constant, $\mu$ is the Lorentz index of the gluon and $i\epsilon$
represents the infinitesimal imaginary part of the propagator that
connects the matrix element to the radiated gluon. The colour labels
of the different particles are denoted by $a,b$ and $c$.  The
representation of the eikonal line is denoted by $R$. We have $R=F$
for the fundamental representation, $R=\bar F$ for the charge
conjugate of the fundamental representation, and $R=A$ for the adjoint
representation.  The colour operators occurring in
Eqs.~(\ref{eq:ineikonaldiagram}) and (\ref{eq:outeikonaldiagram}) are
given in Table~\ref{t:colouroperators}. Note that the order of the
colour indices $a,b,c$ in $f_{abc}$ is kept fixed irrespective of
whether the gluon is emitted above or below the eikonal line.  \TABLE{
\begin{tabular}{ll}
  Outgoing (s)quark / incoming anti-(s)quark: & $(T_F^c)_{ab}=T^c_{ab}$\\
  Outgoing anti-(s)quark / incoming (s)quark: & $(T^c_{\bar F})_{ab}=-T^c_{ba}
  =-(T^c_{ab})^*$\\
  Gluons / gluinos: & $(T^c_A)_{ab}=F^c_{ab}=-if_{abc}$
\end{tabular}
\caption{Colour operators used in the eikonal Feynman rules.
         \label{t:colouroperators}}
}

\section{\boldmath One-loop eikonal integral for $\sq\gl$ production}
\label{sec:appD}

We briefly present here the calculation of the kinematic part
$\omega^{34}$ of the one-loop correction to the process $\,qg \to
\sq\gl\,$ in the eikonal approximation. The equal-mass case of
$\omega^{34}$ is well known \cite{Kidonakis:1997gm}, but for $\sq\gl$
final states we also need the unequal-mass version.

The kinematic part of the one-loop correction generated by the
exchange of a virtual gluon between the two final-state eikonal lines
is according to Eq.~(\ref{eq:outeikonaldiagram}) given by
\begin{equation}
\omega^{34} \;=\; g_s^2 \int \frac{d^d k}{(2 \pi)^d}\,
                  \Bigl(\,\frac{v_3}{v_3 \cdot k + i\epsilon}\,\Bigr)
                  \cdot\Bigl(\frac{v_4}{-v_4 \cdot k + i\epsilon}\,\Bigr)\,
                  \frac{-i}{(k^2+i\epsilon)}\,N^{\mu\nu}(k)\,.
\label{eq:typicaloneloop}
\end{equation}
We use dimensionless vectors $\,v_{i}^{\mu}=p_{i}^{\mu}\sqrt{2/s}\,$
with $p_i$ denoting the momentum of the massive external particle $i$.
We calculate the gluon propagator in a general axial gauge with
\begin{equation}
N^{\mu\nu}(k) \;=\; g^{\mu\nu}-\frac{n^{\mu}k^{\nu}+k^{\mu}n^{\nu}}{n \cdot k}
                    + n^2\frac{k^{\mu}k^{\nu}}{(n \cdot k)^2}\,,
\label{eq:gluonpropagator}
\end{equation}
where $n^{\mu}$ is a general gauge vector with $n^2<0$.  In the case
that $v_{3,4}^2>0\,$ and $\,v_3^2\neq v_4^2$ the solution of the
integral $\omega^{34}$ reads
\begin{equation}
\omega^{34} \;=\; -\,\frac{\alpha_{s}}{\pi\epsilon}\left[ L_{v_3,v_4}  
                  \,+\,L_{v_3} + L_{v_4}\,-1\right]\,,
\label{eq:int_wij}
\end{equation}
with $\epsilon=4-d$. The gauge-independent term $L_{v_3,v_4}$ is given by
\begin{eqnarray}
L_{v_3,v_4} &=& \frac{1}{2}\,\frac{v_3\cdot v_4}
                                  {\sqrt{(v_3\cdot v_4)^2-v_3^2 v_4^2}}\,
\left[\, 2i\pi 
\,+\, \log\left(\frac{v_4^2+v_3\cdot v_4-\sqrt{(v_3\cdot v_4)^2-v_3^2 v_4^2}}
                     {v_4^2+v_3\cdot v_4+\sqrt{(v_3\cdot v_4)^2-v_3^2 v_4^2}}
      \,\right)\right. \nonumber \\[3mm] 
                   && \left.\hphantom{\frac{1}{2}\,\frac{v_3\cdot v_4}
                                     {\sqrt{(v_3\cdot v_4)^2-v_3^2 v_4^2}}a} 
\,+\, \log\left(\frac{v_3^2+v_3\cdot v_4-\sqrt{(v_3\cdot v_4)^2-v_3^2 v_4^2}}
                     {v_3^2+v_3\cdot v_4+\sqrt{(v_3\cdot v_4)^2-v_3^2 v_4^2}}
      \,\right) \,\right]\,.
\label{eq:L_vi_vj}
\end{eqnarray}
The gauge-dependent terms $L_{v_3}$ and $L_{v_4}$ can be found in
Ref.~\cite{Kidonakis:1997gm} and cancel against contributions from the
self-energy diagrams when calculating the anomalous dimensions.  The
gauge-independent term $L_{v_3,v_4}$ can be rewritten in a compact
form using $\,\beta\,$ and $\,\kappa\,$ as defined in
Eqs.~(\ref{eq:logbeta:structure}) and (\ref{eq:kappa}):
\begin{equation}
L_{v_3,v_4} \,=\ \frac{\kappa^2+\beta^2}{2\kappa\beta}\,
                 \biggl[\,\log\Bigl(\,\frac{\kappa-\beta}{\kappa+\beta}\,\Bigr)
                        + i\pi\,\biggr]\,.
\label{eq:Lp3p4} 
\end{equation}
For equal-mass final-state particles this quantity reduces to the
well-known form (cf.~\cite{Kulesza:2009kq})
\begin{equation*}
L_\beta \ =\ \frac{1+\beta^2}{2\beta}\,
             \biggl[\, \log\Bigl(\,\frac{1-\beta}{1+\beta}\,\Bigr) 
                       + i\pi \,\biggr]\,.
\end{equation*}

\end{appendix}


\clearpage
 
\begin{thebibliography}{99}

\bibitem{Golfand:1971iw}
  Yu.~A.~Golfand and E.~P.~Likhtman,
  JETP Lett.\  {\bf 13}, 323 (1971)
  [Pisma Zh.\ Eksp.\ Teor.\ Fiz.\  {\bf 13}, 452 (1971)].

\bibitem{Wess:1974tw}
  J.~Wess and B.~Zumino,
  Nucl.\ Phys.\  B {\bf 70} (1974) 39.

\bibitem{:2007ww}
  V.~M.~Abazov {\it et al.}  [D0 Collaboration],
  Phys.\ Lett.\  B {\bf 660} (2008) 449
  [arXiv:0712.3805 [hep-ex]].

\bibitem{Aaltonen:2008rv}
  T.~Aaltonen {\it et al.}  [CDF Collaboration],
  Phys.\ Rev.\ Lett.\  {\bf 102} (2009) 121801
  [arXiv:0811.2512 [hep-ex]].

\bibitem{Aad:2009wy}
  G.~Aad {\it et al.}  [The ATLAS Collaboration],
  arXiv:0901.0512 [hep-ex].

\bibitem{Bayatian:2006zz}
  G.~L.~Bayatian {\it et al.}  [CMS Collaboration],
  J.\ Phys.\ G {\bf 34} (2007) 995.

\bibitem{gianotti_eps}
F.~Gianotti, talk at the the 2009 Europhysics Conference on High Energy Physics, Krakow, Poland.

\bibitem{Nilles:1983ge}
  H.~P.~Nilles,
  Phys.\ Rept.\  {\bf 110} (1984) 1.

\bibitem{Haber:1984rc}
  H.~E.~Haber and G.~L.~Kane,
  Phys.\ Rept.\  {\bf 117} (1985) 75.

\bibitem{Beenakker:1997ut}
  W.~Beenakker, M.~Kr\"amer, T.~Plehn, M.~Spira and P.~M.~Zerwas,
  Nucl.\ Phys.\  B {\bf 515} (1998) 3
  [arXiv:hep-ph/9710451].

\bibitem{Ellis:1983ed}
  J.~R.~Ellis and S.~Rudaz,
  Phys.\ Lett.\  B {\bf 128} (1983) 248.

\bibitem{Baer:2007ya} 
  see e.g.\ H.~Baer, V.~Barger, G.~Shaughnessy, H.~Summy and
  L.~t.~Wang,
  Phys.\ Rev.\  D {\bf 75} (2007) 095010
  [arXiv:hep-ph/0703289].

\bibitem{Kane:2008kw}
  G.~L.~Kane, A.~A.~Petrov, J.~Shao and L.~T.~Wang,
  arXiv:0805.1397 [hep-ph].

\bibitem{Beenakker:1994an}
  W.~Beenakker, R.~H\"opker, M.~Spira and P.~M.~Zerwas,
  Phys.\ Rev.\ Lett.\  {\bf 74} (1995) 2905
  [arXiv:hep-ph/9412272].

\bibitem{Beenakker:1995fp}
  W.~Beenakker, R.~H\"opker, M.~Spira and P.~M.~Zerwas,
  Z.\ Phys.\  C {\bf 69} (1995) 163
  [arXiv:hep-ph/9505416].

\bibitem{Beenakker:1996ch}
  W.~Beenakker, R.~H\"opker, M.~Spira and P.~M.~Zerwas,
  Nucl.\ Phys.\  B {\bf 492} (1997) 51
  [arXiv:hep-ph/9610490].

\bibitem{Hollik:2007wf}
  W.~Hollik, M.~Kollar and M.~K.~Trenkel,
  JHEP {\bf 0802} (2008) 018.
  [arXiv:0712.0287 [hep-ph]].

\bibitem{Hollik:2008yi}
  W.~Hollik and E.~Mirabella,
  JHEP {\bf 0812} (2008) 087
  [arXiv:0806.1433 [hep-ph]].

\bibitem{Hollik:2008vm}
  W.~Hollik, E.~Mirabella and M.~K.~Trenkel,
  JHEP {\bf 0902} (2009) 002
  [arXiv:0810.1044 [hep-ph]].

\bibitem{Mirabella:2009ap}
 E.~Mirabella,
 arXiv:0908.3318 [hep-ph].

\bibitem{Alan:2007rp}
  A.~T.~Alan, K.~Cankocak and D.~A.~Demir,
  Phys.\ Rev.\  D {\bf 75} (2007) 095002
  [Erratum-ibid.\  D {\bf 76} (2007) 119903]
  [arXiv:hep-ph/0702289].

\bibitem{Bornhauser:2007bf}
  S.~Bornhauser, M.~Drees, H.~K.~Dreiner and J.~S.~Kim,
  Phys.\ Rev.\  D {\bf 76} (2007) 095020
  [arXiv:0709.2544 [hep-ph]].

\bibitem{Kane:1982hw}
  G.~L.~Kane and J.~P.~Leveille,
  Phys.\ Lett.\  B {\bf 112} (1982) 227.

\bibitem{Harrison:1982yi}
  P.~R.~Harrison and C.~H.~Llewellyn Smith,
  Nucl.\ Phys.\  B {\bf 213} (1983) 223
  [Erratum-ibid.\  B {\bf 223} (1983) 542].

\bibitem{Dawson:1983fw}
  S.~Dawson, E.~Eichten and C.~Quigg,
  Phys.\ Rev.\  D {\bf 31} (1985) 1581.

\bibitem{Kulesza:2008jb}
  A.~Kulesza and L.~Motyka,
  Phys.\ Rev.\ Lett.\  {\bf 102} (2009) 111802
  [arXiv:0807.2405 [hep-ph]].
  
\bibitem{Kulesza:2009kq}
  A.~Kulesza and L.~Motyka,
  arXiv:0905.4749 [hep-ph].

\bibitem{Langenfeld:2009eg}
  U.~Langenfeld and S.~O.~Moch,
  arXiv:0901.0802 [hep-ph].

\bibitem{Beneke:2009rj}
  M.~Beneke, P.~Falgari and C.~Schwinn,
  arXiv:0907.1443 [hep-ph].

\bibitem{Beneke:2009nr}
 M.~Beneke, P.~Falgari and C.~Schwinn,
 arXiv:0909.3488 [hep-ph].

\bibitem{Hagiwara:2009hq}
 K.~Hagiwara and H.~Yokoya,
 arXiv:0909.3204 [hep-ph].

\bibitem{Idilbi:2009cc}
  A.~Idilbi, C.~Kim and T.~Mehen,
  Phys.\ Rev.\  D {\bf 79} (2009) 114016
  [arXiv:0903.3668 [hep-ph]].
  
\bibitem{Sdy}
  G.~Sterman,
  Nucl.\ Phys.\  B {\bf 281} (1987) 310.

\bibitem{CTdy}
  S.~Catani and L.~Trentadue,
  Nucl.\ Phys.\  B {\bf 327} (1989) 323.

\bibitem{Contopanagos:1996nh}
  H.~Contopanagos, E.~Laenen and G.~Sterman,
  Nucl.\ Phys.\  B {\bf 484} (1997) 303
  [arXiv:hep-ph/9604313].

\bibitem{Kidonakis:1998bk}
 N.~Kidonakis, G.~Oderda and G.~Sterman,
 Nucl.\ Phys.\  B {\bf 525} (1998) 299
 [arXiv:hep-ph/9801268].

\bibitem{Kidonakis:1998nf} 
  N.~Kidonakis, G.~Oderda and G.~Sterman,
  Nucl.\ Phys.\ B {\bf 531} (1998) 365 [arXiv:hep-ph/9803241].

\bibitem{Bonciani:1998vc}
  R.~Bonciani, S.~Catani, M.~L.~Mangano and P.~Nason,
  Nucl.\ Phys.\  B {\bf 529}, 424 (1998)
  [arXiv:hep-ph/9801375].

\bibitem{Botts:1989kf}
  J.~Botts and G.~Sterman,
  Nucl.\ Phys.\  B {\bf 325} (1989) 62.

\bibitem{Kidonakis:1997gm}
  N.~Kidonakis and G.~Sterman,
  Nucl.\ Phys.\  B {\bf 505} (1997) 321
  [arXiv:hep-ph/9705234].
  
\bibitem{Catani:1996yz}
  S.~Catani, M.~L.~Mangano, P.~Nason and L.~Trentadue,
  Nucl.\ Phys.\  B {\bf 478} (1996) 273
  [arXiv:hep-ph/9604351].


\bibitem{Kulesza:2002rh}
  A.~Kulesza, G.~Sterman and W.~Vogelsang,
  Phys.\ Rev.\  D {\bf 66} (2002) 014011
  [arXiv:hep-ph/0202251].


\bibitem{Martin:2009iq}
  A.~D.~Martin, W.~J.~Stirling, R.~S.~Thorne and G.~Watt,
  Eur.\ Phys.\ J.\  C {\bf 63} (2009) 189
  [arXiv:0901.0002 [hep-ph]].
  
\bibitem{prospino}
see \url{http://www.thphys.uni-heidelberg.de/~plehn/prospino/} or
\url{http://people.web.psi.ch/spira/prospino/}

\end{thebibliography}
\end{document}